\PassOptionsToPackage{colorlinks,citecolor=blue,urlcolor=red}{hyperref}
\PassOptionsToPackage{round,authoryear}{natbib}
\documentclass[sigconf,nonacm,screen]{acmart}

\setcitestyle{round}
\bibpunct{(}{)}{;}{a}{,}{,}

\usepackage{epstopdf}
\usepackage[caption=false]{subfig}
\usepackage{enumitem}
\setlist{leftmargin=5.5mm}
\usepackage[skip=3pt]{caption}
\usepackage{multirow}
\usepackage{array}
\usepackage{orcidlink}
\usepackage[T1]{fontenc}
\usepackage[utf8]{inputenc}
\usepackage{booktabs}
\usepackage{twemojis}
\usepackage[inkscapeformat=pdf]{svg}
\usepackage[hyphenbreaks]{breakurl}
\usepackage[nameinlink]{cleveref}

\begin{document}

\title{Mapping Technological Futures: Anticipatory Discourse Through Text Mining}

\author{Maciej Skórski\orcidlink{0000-0003-2997-7539}}
\affiliation{
  \institution{Czech Technical University Prague}
  \country{Czech Republic}
}
\orcid{0000-0003-2997-7539}
\email{maciej.skorski@gmail.com}

\author{Alina Landowska\orcidlink{0000-0002-7966-8243}}
\affiliation{
  \institution{SWPS University}
  \country{Poland}
}
\orcid{0000-0002-7966-8243}
\email{alandowska@swps.edu.pl}

\author{Krzysztof Rajda}
\affiliation{
  \institution{Brand24}
  \country{Poland}
}
\email{krzysztof.rajda@brand24.com}

\begin{abstract}
The volatility and unpredictability of emerging technologies, such as artificial intelligence (AI), generate significant uncertainty, which is widely discussed on social media. This study examines anticipatory discourse surrounding technological futures by analysing 1.5 million posts from 400 key opinion leaders (KOLs) published on the X platform (2021-2023). Using advanced text mining techniques, including BERTopic modelling, sentiment, emotion, and attitude analyses, the research identifies 100 distinct topics reflecting anticipated tech-driven futures. Our findings emphasize the dual role of KOLs in framing \textit{present futures}—optimistic visions of transformative technologies like AI and IoT—and influencing \textit{future presents}, where these projections shape contemporary societal and geopolitical debates. Positive emotions such as Hope dominate, outweighing Anxiety, particularly in topics like "Machine Learning, Data Science, and Deep Learning," while discussions around "Climate Change" and "War, Ukraine, and Trump People" elicit \textit{Anxiety}. By framing technologies as solutions to societal challenges, KOLs act as mediators of societal narratives, bridging imagined futures and current realities. These insights underscore their pivotal role in directing public attention with emerging technologies during periods of heightened uncertainty, advancing our understanding of anticipatory discourse in technology-mediated contexts.
\end{abstract}


\maketitle

\section{Introduction}
Anticipating future plays a critical role in decision-making and behaviour as the desire to reduce uncertainty is a potent motivator of social behaviour \citep{hirsh_psychological_2012, moore_data-frame_2011, grabenhorst_two_2021}. As people constantly seek to estimate and reduce uncertainties in social interactions to enhance their productivity, well-being, and ultimately their survival as social beings \citep{feldmanhall_resolving_2019}. The ability to navigate and adapt to uncertain futures becomes crucial. Hope, Trust, Fear, and Anxiety are widely regarded as anticipatory emotions \citep{castelfranchi_anticipation_2011, feil_anticipatory_2022}, with \textit{Hope} and \textit{Anxiety} specifically identified as emotions of uncertainty \citep{gordon_emotions_1969, gordon_structure_1987}.

The rise of social media as a hub for technological discourse \citep{koivunen_anticipation_2023} and the pivotal role of digital platforms in structuring anticipatory futures \citep{tavory_coordinating_2013} have drawn significant academic interest. Specifically, research on Twitter has explored various aspects of anticipatory discourse. For instance, studies have examined how different occupational groups perceive generative AI \citep{miyazaki_public_2023}, how public acceptance of emerging technologies like self-driving cars \citep{sadiq_self-driving_2018}, and how Twitter serves as a platform for global discourse, such as during the Paris Climate Talks (COP21) \citep{hopke_visualizing_2018}.

Moreover, prior research highlights the dynamics of user participation on Twitter, particularly the distinction between ordinary users and influential actors who drive discussions \citep{tur_effect_2022, bruns_routledge_2015}. While it is widely recognized that influencers play a significant role in shaping public narratives, the specific ways in which tech influencers shape public expectations of technological advancement remain underexplored. This gap in the literature underscores the need for further investigation into their role in anticipatory technological discourse.

This study examines anticipatory discourse on X\footnote{Formerly known as Twitter, tweets on the platform are now referred to as "posts."}, focusing on the influence of key opinion leaders (KOLs) in shaping public perceptions of technological futures. \Cref{tab:posts_examples} provides examples of anticipatory utterance in the corpus. This study integrates BERTopic modelling, sentiment, and emotion analyses to explore technology-induced futures introduced by X influencers from 2021 to 2023.

\begin{table}[h!]
    \caption{Examples of Anticipatory Discourse of Tech Influencers in the Analysed Corpora} \label{tab:posts_examples}
    \footnotesize
    \begin{tabular}{p{1cm} p{6cm}}
    \toprule
   \textbf{ Topic} & \textbf{Tech Influencers' post example}\\
    \midrule
     9 & Next level of \#remotework - will working from home (\#WFH) be the new standard for the \#futureofwork?\\
     40 & Will you learn to trust artificial intelligence next year?\\
     54 & How is Artificial Intelligence Revolutionizing the Educational Sector?\\
     78 & \#Who’s Next? Security Guards? Yet another job that will be replaced by robots?\\ 
     90 & When will we see the first \#AI generated hit song? Book? \twemoji{thinking face}\\
    \bottomrule
    \end{tabular}  
\end{table}

By employing a triangulated methodological approach (e.g., \cite{egbert_using_2019}), we aim to integrate three distinct methods for discourse analysis of a corpus: BERTopic modelling (contextual corpus analysis), Keywords analysis (non-contextual corpus analysis), and lexicon-based analysis (sentiment, emotion, attitude analyses). Our goal is to identify the leading technologies and KOLs, analyse the evolution and interrelation of technological topics (anticipated futures) over time, and assess the associated sentiments, anticipatory emotions, and attitudes within a corpus of approximately 1.5 million posts from 2021 to 2023. We aim to address the main research question:
"What specific technology-driven futures are being anticipated by KOLs on platform X?" This is explored through the following sub-questions:

\begin{itemize}
\item \textbf{[Rq1]} Who are the KOLs driving these discussions about technological futures, and what are the characteristics of their influence on social media platforms?
\item \textbf{[Rq2]} What are the dominant themes and technologies discussed in anticipatory discourse
\item \textbf{[Rq3]} How do anticipatory discussions about technological futures evolve over time, and what external events or milestones influence these shifts?
\item \textbf{[Rq4]} How are different anticipated futures interconnected, and what patterns of co-occurrence or thematic relationships can be observed in the discourse?
\item \textbf{[Rq5]} To what extent do these discussions align with the concepts of \textit{present future} (visions of the future shaping the present) and \textit{future present} (the present understood through the lens of the future)?
\item \textbf{[Rq6]} What sentiments (positive/negative) and anticipatory emotions (e.g., \textit{Hope}, fear, excitement) are expressed by KOLs towards specific technological futures?
\end{itemize}

While exploratory, our research questions align with the theoretical foundations outlined by \cite{tavory_coordinating_2013}, which advocate for studying how societies coordinate their futures through anticipatory interactions.

\section{Literature Review}

Research on how individuals anticipate and structure future events has been central in social theory, e.g., \citep{adam_time_1990, bergmann_problem_1992, emirbayer_what_1998, abbott_time_2001, mische_projects_2009}. Every human interaction inherently involves a connection to future events \cite{tavory_coordinating_2013}. When individuals engage with each other, they orient themselves towards future circumstances, and consequently it becomes imperative for individuals to simultaneously navigate the inherent uncertainties associated with these future-oriented interactions(\textit{ibid.}, p.~909). 

\textbf{Anticipating Futures.} We recognize two main categories of the technological future: a. \textit{present futures} \citep{luhmann_differentiation_1982} are ‘pre-given futures’ rooted in the past, i.e., lived ones \citep{adam_futures_2011}; and b. \textit{future presents} \citep{luhmann_differentiation_1982} are "futures-in-the-making" that are possibly latent, growing, and changing ones, i.e., living ones \citep{adam_futures_2011}. To clarify these concepts, we provide an expanded distinction (\Cref{tab:future_distinction}).

\textit{Present futures} are linear continuations of the past into the present \citep{poli_anticipation_2014}, grounded in historical continuity and representing a concretized future. They are practical, explicit, contextualized, embodied, and embedded \citep{adam_future_2007}. These futures shape the present by present means \citep{adam_future_2007, miller_futures_2007}, and their value is calculated against alternatives (e.g., forecasts) and traded as commodities \citep{adam_future_2007}. The deterministic and calculable nature of these futures evokes emotions such as \textit{Hope} or fear, serving to mobilize action \citep{luhmann_differentiation_1982}.

In contrast, \textit{future presents} are latent possibilities that can be recognized and foreseen, influencing the present by entering into it \citep{adam_futures_2011}. These are abstract and symbolic, removed from immediate contexts and open to exploration \citep{adam_future_2007}. As open-type futures, they embrace high uncertainty \citep{beckert_capitalism_2013, poli_anticipation_2014}, evoking fictional expectations based on the “as if” rule \citep{beckert_capitalism_2013}. These futures inspire novel reconfigurations of the present and are anticipatory in their orientation, focusing on the `use-of-the-future` \citep{poli_introduction_2017}.

\begin{table}[t]
\centering
\caption{Distinction Between \textit{Present Futures} and \textit{Future Presents}}
\resizebox{0.99\linewidth}{!}{
\begin{tabular}{p{2.2cm} p{3.1cm} p{3.1cm}}
\hline
\textbf{Aspect} & \textbf{Present Futures} & \textbf{Future Presents} \\ \hline
\textbf{Nature} & Concrete, explicit, contextualized \citep{adam_future_2007} & Abstract, implicit, decontextualized \citep{adam_future_2007} \\ 
\textbf{Temporal Orientation} & Rooted in the past, extended into the future \citep{poli_anticipation_2014} & Latent possibilities emerging in the present \citep{adam_futures_2011} \\
\textbf{Predictability} & Linear, calculable \citep{adam_future_2007} & Open-ended, unpredictable \citep{beckert_capitalism_2013} \\
\textbf{Role} & Guides immediate action \citep{adam_future_2007} & Inspires imagination and long-term vision \citep{poli_introduction_2017} \\ 
\textbf{Value} & Practical, tangible \citep{adam_future_2007} & Symbolic, theoretical \citep{adam_future_2007} \\ 
\textbf{Example} & Market forecasts, urban plans \citep{poli_anticipation_2014} & Science fiction, speculative futures \citep{beckert_capitalism_2013} \\ \hline
\end{tabular}
}
\label{tab:future_distinction}
\end{table}

By emphasizing these distinctions, we aim to clarify how the  \textit{present futures} serve as imagined or projected from the current moment (e.g., hopes, expectations, visions), while \textit{future presents} inspire innovative thinking by embracing uncertainty and speculative exploration (e.g., how current actions, debates, or technologies are influenced by visions of the future).

\textbf{Anticipatory Emotions.} Anticipation plays a crucial role in the generation of emotions, and the literature distinguishes between future-oriented emotions for anticipatory and anticipated experiences. \textit{Anticipatory emotions} are present feelings tied to an upcoming event \citep{bagozzi_goal-directed_1998, baumgartner_future-oriented_2008, feil_anticipatory_2022}, while \textit{anticipated emotions} relate to the expected emotional responses a person may have when a future occurrence comes to pass \citep{baumgartner_future-oriented_2008, feil_anticipatory_2022, perugini_role_2001}. Anticipatory emotions typically involve prospective feelings such as \textit{Hope} and Fear, whereas anticipated emotions are more retrospective, encompassing sensations like Relief, Satisfaction, Disappointment, and Anger \citep{baumgartner_future-oriented_2008, feil_anticipatory_2022}. For instance, \textit{Hope} is associated with positive potential outcomes, while Fear is tied to possible negative ones \citep{castelfranchi_anticipation_2011, macleod_anticipatory_2017, vazard_feeling_2024}. Disappointment arises when \textit{Hope} is unmet, while Relief follows the avoidance of a Feared event \citep{baumgartner_future-oriented_2008, mowrer_learning_1960}.

Plutchik’s psychoevolutionary theory of emotions \citep{plutchik_general_1980, plutchik_nature_2001} provides a detailed framework that explains how Anticipation interacts with other primary emotions. Plutchik identifies eight primary bipolar emotions, including Anticipation and its opposite, Surprise. These emotions are visualized as a circumplex model, often depicted as a flower with "petals" representing each emotion and their interrelationships.

The combinations of Anticipation with other primary emotions form distinct emotional dyads, which Plutchik categorizes as primary, secondary, or tertiary based on their psychological closeness:

\begin{itemize}
    \item Primary dyads (adjacent petals) involve Anticipation combining with Joy to create \textit{Optimism}, or with Anger to produce \textit{Aggressiveness}.
    \item Secondary dyads (two petals apart) involve Anticipation paired with Trust to generate \textit{Hope}, or with Disgust to evoke \textit{Cynicism}.
    \item Tertiary dyads (three petals apart) combine Anticipation with Sadness to result in \textit{Pessimism}, or with Fear to lead to \textit{Anxiety}.
\end{itemize}

Through this lens, the concept of "petals apart" refers to the relative proximity of emotions in the circumplex, with closer emotions being more psychologically interconnected. For instance, \textit{Hope} (Anticipation and Trust) is more closely related to Anticipation than Pessimism (Anticipation and Sadness). This model provides a structured way to analyse how anticipation influences emotional responses and how it interacts with broader affective states, grounding these constructs within established theories of emotion.

These classifications demonstrate the nuanced interactions between Anticipation and other emotions, emphasizing how anticipation bridges both positive (e.g., Optimism, \textit{Hope}) and negative (e.g., \textit{Anxiety, Cynicism}) experiences. The analysis of anticipatory emotions, such as \textit{Hope} and \textit{Anxiety}, builds on the foundational work of \cite{plutchik_general_1980, plutchik_nature_2001} and \cite{castelfranchi_anticipation_2011}, highlighting the emotions expressed by tech influencers in responses to uncertain technological scenarios.

\section{Data and Methods}\label{sec:methodology}

\subsection{Data Collection and Preprocessing} The dataset was collected using the scraping library \texttt{snsscrape} ~\citep{justanotherarchivist_snscrape_2023}. The data were primarily restricted to posts written in English. This decision was made to ensure consistency in the analysis, given the limitations of NLP tools, which perform more reliably on English-language data. While the dataset likely includes contributions from users across the globe, it is inherently biased toward English-speaking users and regions where English is the dominant or widely used language. posts were sourced from about 400 technology influencers` feeds published in the time frame from January 1, 2021, to March 31, 2023. The influencers were selected by a multidisciplinary team of experts comprising (social media analyst, technology domain expert, and data scientist). The selection process was guided by clearly defined criteria, including domain expertise, consistency in engaging with technology-related content, and their active participation in discussions on social media platforms. Additionally, to ensure objectivity, we cross-referenced the selected KOLs with external databases and published lists of influential technology figures on social media. \Cref{tab:dataset_stats} provides a comprehensive overview of the size and diversity of our dataset.

\begin{table}[]
    \caption{Dataset summary statistics}
    \label{tab:dataset_stats}
    \begin{tabular}{p{5cm} p{2cm}}
    \toprule
     Unique accounts & 400 \\ 
     Unique texts & 1,200,003 \\  
     Unique posts (timestamp+text) & 1,458,018 \\  
    \bottomrule
    \end{tabular}
\end{table}

To ensure the accuracy of this analysis, we preprocessed the posts by removing URLs, email addresses, and X user handles.

\subsection{Data Analysis} We employed the triangulated methodology combining BERTopic modelling, sentiment analysis, and emotion analysis. Appropriate non-parametric statistical tests were used to assess the significance of findings. To derive meaningful insights, the outputs of the analytical methods were systematically integrated: a. The topics identified through BERTopic modelling served as the primary framework for categorizing discourse; b. Sentiment and emotion analyses were applied to posts within each topic to assess the dominant sentiments and emotional tones; c. The outputs were combined by mapping sentiment and emotion scores to the corresponding topics, enabling us to identify patterns such as the co-occurrence of emotions (e.g., anticipation and fear) within specific thematic areas. This integrative approach allowed us to analyse not only the quantitative relationships between the methods but also to validate our findings qualitatively by exploring specific themes and contextual nuances.

\textit{Contextual Corpus Analysis.} Following prior work on extracting topics from Twitter data \citep{landowska_what_2023, egger_topic_2022, yang_standing_2021}, we trained the state-of-the-art BERTopic model implemented in the \texttt{BERTopic} library \citep{grootendorst2022bertopic}. Due to the model's vulnerability to data volume, we used online learning with specifications detailed in \Cref{tab:bertopic_specs}. Applied to the entire dataset, BERTopic categorized it into 100 contextually coherent, distinct topics.

\begin{table}[h]
\centering
\caption{BERTopic Model Specifications}
\label{tab:bertopic_specs}
\resizebox{0.99\linewidth}{!}{
\begin{tabular}{ll}
\hline
\textbf{Component} & \textbf{Specification} \\
\hline
Text Model & \texttt{all-MiniLM-L6-v2} \\
Dimensionality Reduction & \texttt{IncrementalPCA} \\
Clustering & \texttt{MiniBatchKMeans} \\
Vectorization & \texttt{OnlineCountVectorizer} \\
Learning Method & Online, with batch size 100,000 \\
Number of Topics & 100 \\
\hline
\end{tabular}
}
\end{table}

To determine the optimal number of topics, we conducted extensive experimentation with various clustering configurations. We evaluated topic coherence scores across different numbers of topics (25-150) and performed hierarchical cluster agglomeration at various thresholds. Human evaluators inspected document samples from each topic to assess semantic coherence. This systematic evaluation revealed that 100 topics provided the optimal balance between diversity and coherence, effectively capturing distinct thematic clusters while maintaining meaningful separation between topics. We visualised the topics using the method of \citep{sievert_ldavis_2014}, implemented in the library \texttt{pyLDAvis}~\citep{mabey_pyldavis_2021}.

In addition to quantitative coherence analysis, we manually inspected the top indicative words for each topic. This qualitative evaluation ensured the interpretability of the topics, enabling us to assign meaningful names based on the underlying themes. For instance, topics such as Leadership of Change and Quantum Computing were named after careful analysis of their most representative terms and context in the dataset.

\textit{Non-contextual Corpus Analysis.} 
Our analysis of the non-contextual corpus consists of identifying KOLs and keywords. KOLs in tech discussions were identified based on their post volume on the X platform \citep{casalo_influencers_2020}. Indicating keywords required utilising a compact English model from the  Spacy\footnote{\url{https://spacy.io/}} package. This process included lemmatization and the exclusion of stop words, resulting in a cleaner and more focused dataset. The remaining lemmatized words formed the basis for keyword extraction. Distinctive keywords were determined by comparing the term frequencies across topics using the class-adjusted Term Frequency-Inverse Document Frequency (c-TF-IDF) weighting. This approach ensured that words that appeared frequently in one topic but were less common across the corpus were identified as distinctive. The analysis of these keywords helped to validate and enrich our topic modelling results.

\textit{Sentiment and Emotions Analysis.} To analyse sentiment, we leveraged the Brand24 model \citep{augustyniak_massively_2023, rajda_assessment_2022}. This recent model is based on a fine-tuned multilingual LLM and is specifically tailored for analysing social media data. The Brand24 sentiment classifier operates in three polarities (positive, negative, and neutral). Refer to \Cref{eq:sentiment} for calculating the average sentiment score. 
\begin{align}\label{eq:sentiment}
\mathrm{Sentiment\ Score_{Avg}}=\frac{N^{+}-N^{-}}{N^{-}+N^0+N^{+}}
\end{align}
where, $N^{-},N^0$ and $N^{+}$ denote the number of posts classified as positive, neutral, and negative, respectively. For emotion analysis, we utilized the Cardiff NLP emotion detection model \citep{mohammad_semeval-2018_2018, camacho-collados_postnlp_2022}, which classifies posts into 11 emotion categories (sadness, disgust, anger, fear, anticipation, surprise, joy, love, trust, pessimism, optimism). We used the official implementation available from the Hugging Face repository\footnote{\url{https://huggingface.co/cardiffnlp/twitter-roberta-base-emotion-multilabel-latest}}. 
Pessimism and optimism scores were rejected because these are not emotions. Instead, we propose a coherent approach by interpreting them as a manifestation of a 'can-do' attitude \cite{bortolotti_optimism_2018, jefferson_what_2017}. We suggest, according to \cite{plutchik_nature_2001} and \cite{tenhouten_emotions_2023}, to define optimism as the cumulative score of anticipation and joy, while pessimism as the combined score of anticipation and sadness. \Cref{eq:secondary_attitude} explains this approach that provides a more structured and verified method to measure these attitudes.
\begin{align}\label{eq:secondary_attitude}
\begin{aligned}
\mathrm{S_{Opt}} &= \mathrm{S_{Ant}} + \mathrm{S_{J}}\\
\mathrm{S_{Pes}} &= \mathrm{S_{Ant}} + \mathrm{S_{S}},
\end{aligned}
\end{align}
where $\mathrm{S_{Opt}}$ denotes the Optimism score, $\mathrm{S_{Ant}}$ denotes the Anticipation score, $\mathrm{S_{J}}$ denotes the Joy score, $\mathrm{S_{Pes}}$ denotes the Pessimism score, and $\mathrm{S_{S}}$ denotes the Sadness score.
Following \cite{plutchik_general_1980}, to define scores of Hope and Anxiety, for each topic, we paired anticipation with trust to obtain Hope scores, while also linking anticipation with fear, resulting in Anxiety (refer \Cref{eq:secondary_emotions}).
\begin{align}\label{eq:secondary_emotions}
\begin{aligned}
\mathrm{S_{Anx}} &= \mathrm{S_{Ant}} + \mathrm{S_{F}}\\
\mathrm{S_{H}} &= \mathrm{S_{Ant}} + \mathrm{S_{T}},
\end{aligned}
\end{align}
where $\mathrm{S_{Anx}}$ denotes the Anxiety score, $\mathrm{S_{Ant}}$ denotes the Anticipation score, $\mathrm{S_{F}}$ denotes the Fear score, $\mathrm{S_{H}}$ denotes the Hope score, and $\mathrm{S_{T}}$ denotes the Trust score and the topics are scored by averaging, respectively. 
We incorporated an emotion colour palette inspired by Plutchik's wheel of emotions from \cite{semeraro_pyplutchik_2021} to visually enhance our findings with an intuitive representation of emotional states. 

\subsection{Statistical Analysis}
Spearman’s rank correlation was selected for statistical analysis due to the ordinal nature of the emotion scores and the lack of normal distribution in the dataset. Spearman’s correlation is well-suited for analysing monotonic relationships without assuming linearity, making it ideal for emotion and sentiment data \citep{zar_biostatistical_1999}. To account for multiple comparisons and reduce the risk of Type I errors, we applied the Bonferroni Correction when testing the significance of correlations between emotion pairs. The Bonferroni method adjusts the \(p\)-value threshold by dividing the significance level (\(\alpha = 0.05\)) by the number of tests conducted, ensuring stringent control of the family-wise error rate \citep{haynes_bonferroni_2013}.

\section{Results}
\subsection{Modelled Topics}

\paragraph{Distinct Topics.}
Based on the optimization outlined, BERTopic was trained to identify 100 topics. Each topic is associated with a specific set of posts, allowing for a deeper understanding of the main themes within the dataset. For a comprehensive list of all topics, refer to the supplementary repository. The Coherence Value measure (refer to \citep{roder_exploring_2015}) for 100 topics is about 66\%, demonstrating that topics are of high-quality (as evidenced by other studies on X data, under this metric, BERTopic rarely achieves scores higher than 70\% \citep{campagnolo_topic_2022,santakij_analyzing_2024,austin_uncovering_2024} unless for small and coherent corpora \citep{chen_leveraging_2023}.

\begin{figure}
\centering\includesvg[width=0.99\columnwidth,inkscapelatex=false,extractformat=eps,convertformat=png]{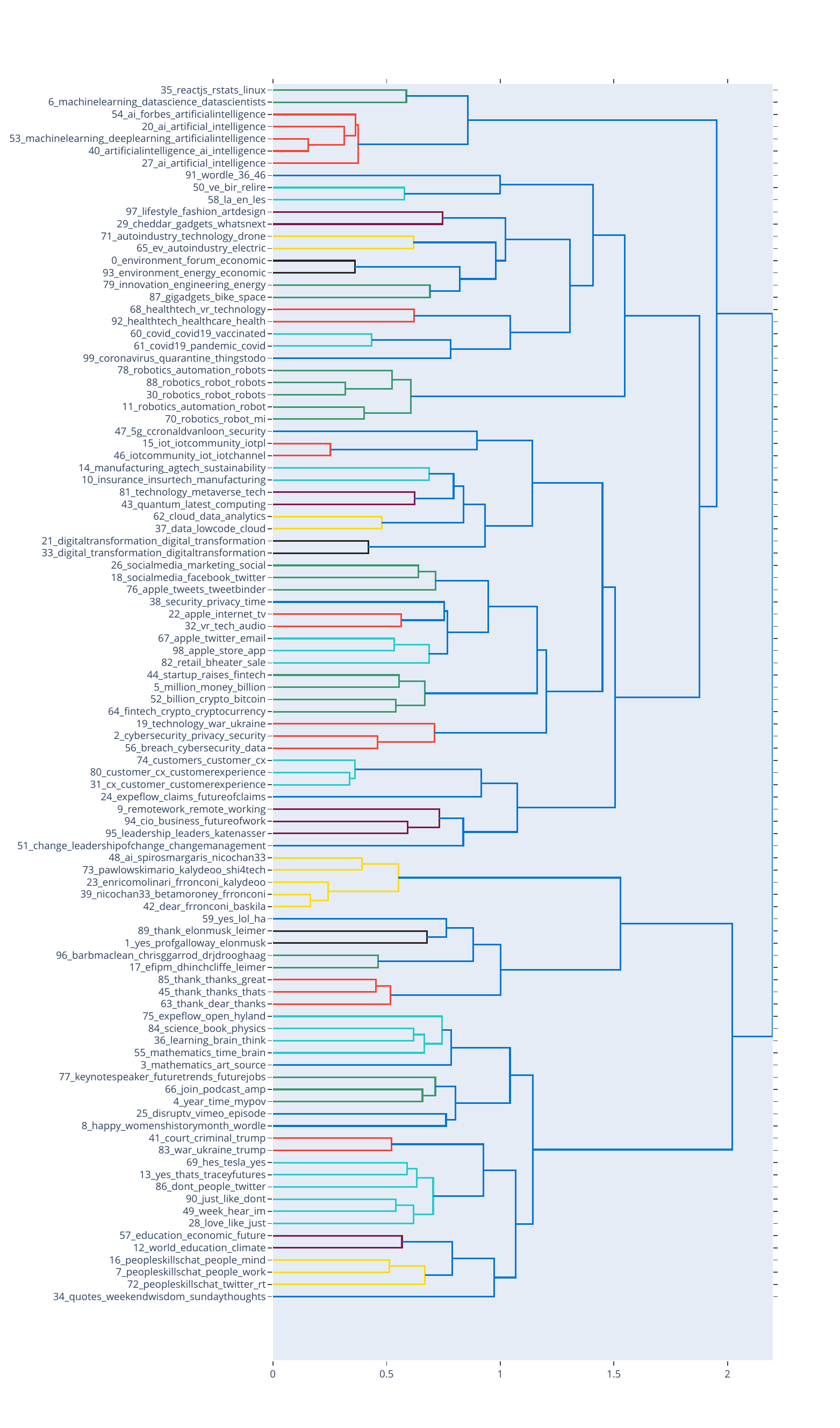}
    \caption{The Hierarchical Clustering of Anticipated Technology-Driven Futures}
    \label{fig:clusters}
\end{figure}

\paragraph{Hierarchical Clustering.} \Cref{tab:BERT_clusters} presents 21 unique clusters generated from 100 topics using BERTopic model. The relationships between the documents and these themes are elucidated, as depicted in the hierarchical clustering dendrogram (refer to \Cref{fig:clusters}). These groups show how different tech themes like AI, ML, and Data Science are influencing each other as they develop within the discussing community. The dendrogram illustrates how topics are related to each other based on their distance in the high-dimensional space, where topics are closer if they are more similar. Additionally, the Intertopic Distance Map (available in the supplementary repository) illustrates how topics are distributed within the semantic space of the corpus. It visually represents the proximity and relationships between topics, with closely clustered topics indicating thematic overlap, while more distant topics suggest greater differentiation in content. For example, the alignment of "AI, Artificial Intelligence, and Machine Learning" (Topic 27),  "VR, Tech, Audio, and AR" (Topic 32), and "AI, Artificial Intelligence, and Big Data" (Topic 20) highlights a network of mutually influential themes. While "Space exploration, Satellites, Drones, and Advanced Engineering Technologies" (Topic 5) intersects with "Metaverse, VR, AR" (Topic 34), "Futurists, Future, Discussion" (Topic 14).

\paragraph{Engaging discussions.} The range of post counts across topics is 23,144 posts (from 7,633 to 30,777). The average engagement of discussion (the mean count of posts per topic) is 14,580 posts. Approximately 41.75\% of the topics exceed this mean count. After excluding non-technology related discussions like "Time Reflections" or "Gratitude," the top 5 tech-driven topics include "Disruptv, Vimeo, and Episode Video" (Topic 25) with 27,231 posts, "Climate Change" (Topic 0) with 27,103 posts, "War, Ukraine, and Trump People" (Topic 83) with 22,485 posts, "Machine Learning, Data Science, and Deep Learning" (Topic 6) with 21,827 posts, and "IoT Community and IoT Channel" (Topic 46) with 21,635 posts. These topics are discussed significantly more frequently than the bottom 5 below the median, such as "Robotics, Robot, and Ronald Van Loon" (Topic 70) with 7,633 posts, "Coronavirus, Quarantine, and Things to Do" (Topic 99) with 8,418 posts, "HealthTech, VR, and Brain Technology" (Topic 68) with 8,430 posts, "Innovation and Engineering in Technology" (Topic 79) with 8,792 posts, and "Manufacturing, Agtech, and Sustainability Innovation" (Topic 14) with 9,096 posts.

\begin{figure}[h!]
\centering
\includegraphics[width=0.99\columnwidth]{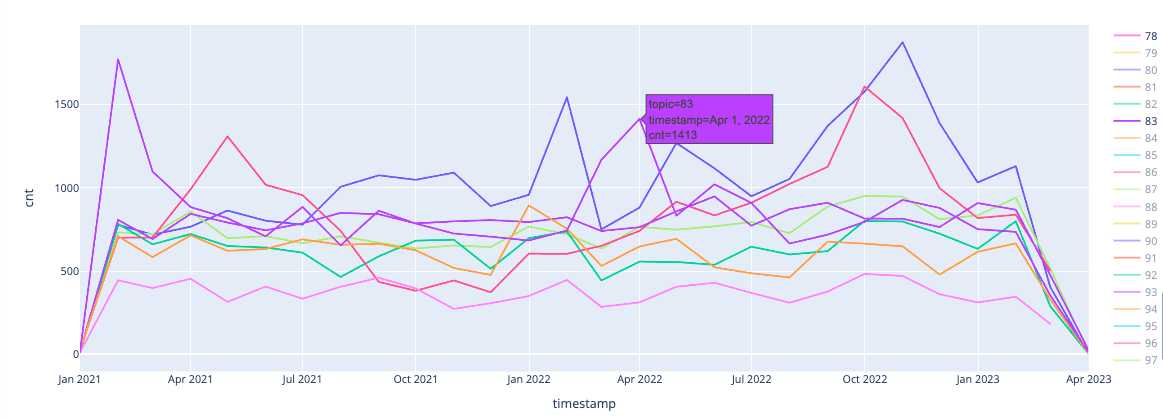}
\caption{The Dynamics of Selected Topics over Time}
\label{fig:topic-evolution}
\end{figure}

\paragraph{Temporal Dynamics.} All topics demonstrate a consistently ongoing and stable discussion over time, without significant declines or interruptions (refer to \Cref{fig:topic-evolution}). Unlike other themes that might experience sharp spikes followed by a drop in interest, this discourse remains sustained and resilient. Its continuous presence suggests a long-term relevance, where time does not diminish engagement or the importance of the discussion. For instance, "Deep Learning and Data Science" (Topic 53). In contrast, other topics demonstrate increases or decreases at specific points. For instance, "Climate Change" (Topic 0) (the blue line) reached an all-time high with two significant peaks. The first peak occurred during the World Economic Forum (WEF)  from 17 to 21 January 2022, and the second during the 27th United Nation Conference of the Parties (COP) 27 held from 6 November until 20 November 2022. "IoT Community and IoT Channel" (Topic 46) (the red line) showed a peak around the Special Annual Meeting of the WEF from May 25-28, 2021. Some topics, such as "War, Ukraine, and Trump People" (Topic 83) (the violet line), manifest notable fluctuations, while others like "Robotics, Automation, and Autonomous Robots" (Topic 78) (the pink line) remain relatively stable until they drop off at the end. For instance, "AI, Artificial Intelligence, and Machine Learning" (Topic 27) (the light green line), "VR, Tech, Audio, and AR" (Topic 32) (the dark green line), and "AI, Forbes, and Artificial Intelligence" (Topic 54) (the orange line) seems to resonate with each other.

\begin{table}[h!]
    \caption{The organisation of Topics into Clusters by BERTopic Analysis}
    \label{tab:BERT_clusters}
    \begin{tabular}{p{1.0cm} p{1.9cm} p{4.1cm}}
    \toprule
    \textbf{Cluster} & \textbf{Topics} & \textbf{Leitmotiv}\\
    \midrule
    1 & 35, 6, 54, 20, 53, 40, 27 & AI, ML, DL, Big Data, NLP\\ 
    2 & 91, 50, 58 & Entertainment and Gaming\\  
    3 & 97, 29 & Fashion and Lifestyle\\ 
    4 & 71, 65, 0, 93, 79, 87 & AV, Automatic Industry, EV, Eco-energy, Climate Change, Innovative engineering, Drones, Space \\ 
    5 & 68, 92 & Healthcare, AI\\ 
    6 & 60, 61, 99 & Covid-19 and Quarantine\\ 
    7 & 78, 88, 30, 11, 70 & Automation, Robots Engineering, Robotics, Autonomous Robots, AI\\ 
    8 & 47, 15, 46 & IoT, 5G, Security, Digital Transformation, Industry 4.0, Edge Computing\\
    9 & 14, 10, 81, 43, 62, 37, 21, 33 & Quantum Computing, Data Analytics, Digital Transformation, Tech Innovations,  Metaverse\\ 
    10 & 26, 18, 76, 38, 22, 32, 67, 98, 82 & Women in Tech, Social Media, Payments, Retail, Privacy, Security\\
    11 & 44, 5, 52, 64 & Start-up, Fintech, Cyrptocurrency\\
    12 &  19, 2, 56 & Cybersecurity, War, Ukraine\\
    13 &  74, 80, 31, 24 & CX, UX, Future of Claims, Insurtech\\
    14 &  9, 94, 95, 51 & Future of Work, Future of Leaders, Leadership of Change, Strategy, Management\\
    15 &  48, 73, 23, 39, 42 & AI, Data Science, AV, VR, Digital Transformation, Metaverse\\
    16 &  59, 89, 1, 96, 17 & Gratitude (Thanks, Acknowledgements\\
    17 &  85, 45, 63 & Appreciation (Thanks, Great)\\
    18 &  75, 84, 36, 55, 3 & Mathematics, Life, Questioning\\ 
    19 &  77, 66, 4, 25, 8 & Time Reflections (Year, Mypov, Women in History)\\ 
    20 &  41, 83, 69, 13, 86, 90, 49, 28 & Politics, Policies, Ukraine War, Social Media\\ 
    21 &  57, 12, 16, 7, 72, 34 & Future, Economics, Education, People Skills, Collaborative Enterprise, Climate, Teamwork\\ 
    \bottomrule
    \end{tabular}
\end{table}

\subsection{Key Influencers and Technologies}
\begin{figure}
    \centering
    \includegraphics[width=0.99\linewidth]{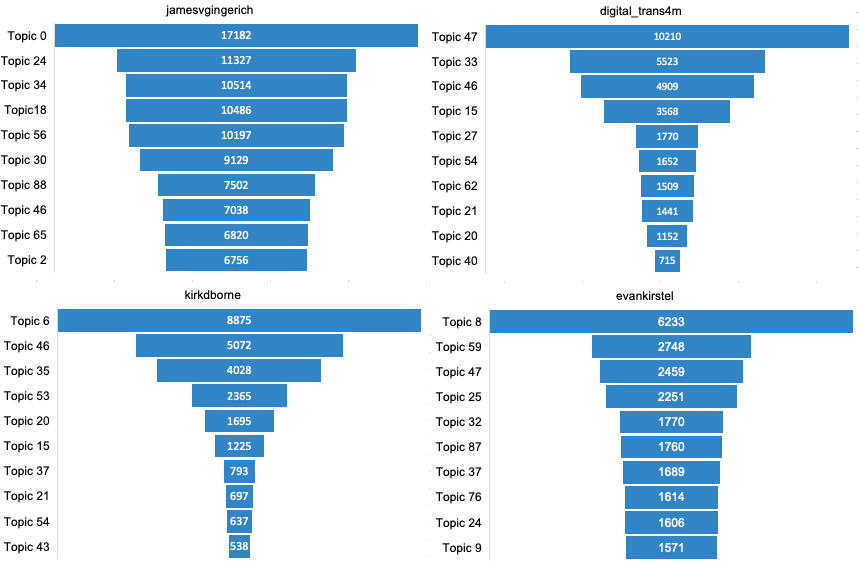}
    \caption{Number of posts by Top-Performing KOLs in the Corpus}
    \label{fig:selected_performers}
\end{figure}

\paragraph{Leading performers.} The top 10 KOLs, who are subject matter experts making significant contributions to specific topics, generated 13\% (187,645 posts) of the total posts. 
They play a pivotal role in steering discussions across 22 topics, dominating these conversations with a substantial share of over 49\%. The data clearly shows distinct specialisation among the KOLs, except for jamesvgingerich, who is active in 46 topics. \Cref{fig:selected_performers} showcases the top performers jamesvgingerich is the standout performer, leading in 10 topics. Among the other key performers, digital\_trans4m is the leading contributor in "5G, Ronald Van Loon, and Security Network" (Topic 47), while kirkdborne leads in "Machine Learning, Data Science, and Deep Learning" (Topic 6), and evankirstel leads in "Happy, Women's History Month, and Wordle Year" (Topic 8). However, these users do not hold leading positions in other topics. While most topics are evenly distributed among users (e.g., "VR, Tech, Audio, and AR" (Topic 32)), "Robotics, Robot, and Engineering" (Topic 88) stands out as an exception. Here, user jamesvgingerich has a significant lead responsible for a staggering 70\% of all posts on this topic, amounting to 7502 posts. In comparison, the second-highest contributor, ronald\_vanloon, has made 951 posts, which constitutes 9\% of the total contributions to this topic.

To illustrate how the two frameworks, \textit{present futures} and \textit{future presents}, are reflected in the analyzed KOL discourse, \Cref{tab:future_discourse} presents examples of posts categorized under each framework. These examples demonstrate how KOLs navigate between projecting tangible visions of transformative technologies and engaging with speculative, long-term debates on societal implications.

\begin{table}[h]
\caption{Examples of \textit{Present Futures} and \textit{Future Presents} in KOLs' Discourse}
\label{tab:future_discourse}
\renewcommand{\arraystretch}{1.3}
\resizebox{0.99\linewidth}{!}{
\begin{tabular}{p{0.19\linewidth}  p{0.8\linewidth}}
\toprule
\textbf{Category} & \textbf{Examples} \\
\midrule
\multirow[b]{3}{5em}{\textit{Present Futures} }     &  Next level of \#remotework - will working from home (\#WFH) be the new standard for the \#futureofwork?  \\
&  How is Artificial Intelligence Revolutionizing the Educational Sector?   \\
&  When will we see the first \#AI-generated hit song?
\\ \midrule
\multirow[b]{3}{5em}{\textit{Future Presents}}    & 
Will you learn to trust artificial intelligence next year?   \\
& \#Who’s Next? Security Guards? Yet another job that  will be replaced by robots?  \\
& What ethical dilemmas will emerge as AI gains  decision-making power?  \\
\bottomrule
\end{tabular}
}
\end{table}

\begin{figure}
    \centering
\includesvg[width=0.99\linewidth,inkscapelatex=false]{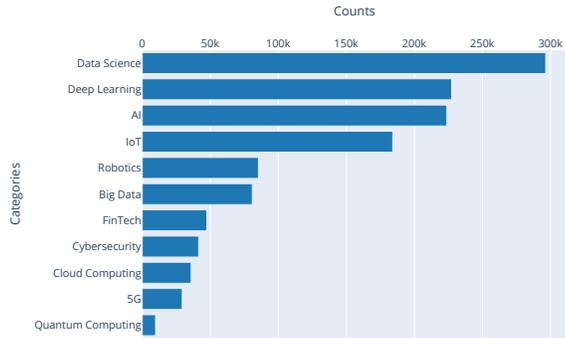}
    \caption{Leading Technologies in the Corpus}
    \label{fig:top_technologies}
\end{figure}

\paragraph{Leading technologies.}
\Cref{tab:repres_keywords} highlights the top 5 "technological" topics, along with their associated representative keywords, providing insight into the core areas of focus within the dataset. While these keywords provide simplified markers, they enable the identification of leading technologies extensively discussed among KOLs (refer to \Cref{fig:top_technologies}).

\begin{table}[]
\caption{The Most Representative Keywords in Selected
Topics}
\label{tab:repres_keywords}
\begin{tabular}{p{0.12\linewidth} p{0.77\linewidth}}
\toprule
\textbf{Topic} & \textbf{Related Keywords}\\
\midrule
2 & ’cybersecurity’, ’privacy’, ’security’,
’cyber’, ’cyberattack’, ’cyberwar’, ’cybercrime’,
’hackers’, ’data’, ’ukraine’\\  
40 & ’artificialintelligence’, ’ai’, ’intelligence’,
’artificial’, ’nlp’, ’machinelearning’,
’robots’, ’deeplearning’, ’read’,
’bigdata’ \\ 
43 & ’quantum’, ’latest’, ’computing’,
’trends’, ’cloud’, ’quantumcomputing’,
’tech’ \\ 
79 & ’innovation’, ’engineering’, ’energy’,
’technology’, ’scientists’, ’interesting’,
’future’, ’space’, ’science’, ’world’\\ 
\bottomrule
\end{tabular}
\end{table}

\subsection{Sentiment, Emotional Patterns, and Attitudes in Anticipatory Discourse}

\paragraph{Sentiment Score.} The sentiment analysis of the dataset reveals that the majority of posts (68.9\%)  are classified as neutral, followed by positive (21.5\%) and negative (9.6\%). 
The average sentiment score ranges from $-$0.5038 (very negative) to 0.8419 (very positive), with a mean of approximately 0.116, indicating an overall positive sentiment across all topics. This cnclusion is further supported by the statistically significant result of the one-sided Wilcoxon signed-rank test ($p=2\cdot 10^{-18})$, which compares positive and negative sentiment across topics.
Topics such as "Climate Change" (Topic 0), "Machine Learning, Data Science, and Deep Learning" (Topic 6), "Disruptv, Vimeo, and Episode Video" (Topic 25), "IoT Community and IoT Channel" (Topic 46), "Artificial Intelligence and Intelligence Artificial" (Topic 40), "Digital Transformation and CCDX Latest" (Topic 33), "AI, Artificial Intelligence, and Machine Learning" (Topic 27), "Expeflow, Claims, and Future of Claims Workflow" (Topic 24), "5G, Ronald Van Loon, and Security Network" (Topic 47), and "AI, Artificial Intelligence, and Big Data" (Topic 20) represent the highest neutral scores. The most positive sentiment dominates in topics such as "Yes, LOL, and Ha" (Topic 59), "Thank, Elon Musk, and Leimer Thanks" (Topic 89), "Yes, Prof Galloway, and Elon Musk Thanks" (Topic 1), "Barb MacLean, Chris G Garrod, and Dr JD Rooghaag" (Topic 96), "EFIPM, Dhinchcliffe, and Leimer CTO Advisor" (Topic 17), "Thank, Thanks, Great, and Happy" (Topic 85), "Thank, Thanks, and That's Love" (Topic 45), and "Thank, Dear, Thanks, and Congratulations" (Topic 63). The most negative sentiment is associated with topics such as "War, Ukraine, and Trump People" (Topic 83), "Court, Criminal, Trump, and Police" (Topic 41), "Social Media, Facebook, Twitter, and Politics" (Topic 18), "Don't, People, Twitter, and Think" (Topic 86), "Cybersecurity, Privacy, Security, and Cyber" (Topic 2), and "Breach, Cybersecurity, Data, and Privacy" (Topic 56).

\begin{figure}
    \centering
\includesvg[width=1.0\linewidth,inkscapelatex=false]{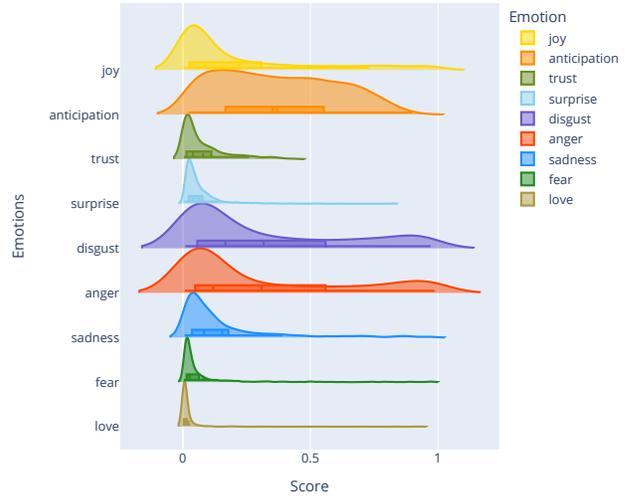}
\caption{Distribution of Emotions across the Corpus}
\label{fig:emotions_descriptive}
\end{figure}

\paragraph{Emotional Patterns.} The distribution of emotions is presented in \Cref{fig:emotions_descriptive}, for more numerical details refer to the OSF repository. \Cref{tab:emotion_scores} provides the 5 topics with the highest emotion scores (above the median) and the 5 topics with the lowest emotion scores (below the median). By employing the non-parametric Spearman’s rank correlation coefficient, even after applying the Bonferroni Correction \(p\)-values ($p=10^{-16}$, we observed significant correlations across all emotion pairs: 
i. The correlation between Anticipation and Trust is 0.455, indicating that Anticipation enhances Trust; ii. Higher Anticipation is associated with lower Sadness, as indicated by a moderate negative correlation of -0.353; iii. There is a weak positive correlation of 0.217 between Anticipation and Fear suggesting that as Anticipation increases, Fear tends to rise slightly; iv. The very weak correlation of 0.007 suggests that Anticipation and Joy are largely independent. When Anticipation increases, Joy does not necessarily change significantly.

\begin{table}[]
    \caption{Highest and Lowest Emotion Scores in Topics}
    \label{tab:emotion_scores}
    \begin{tabular}{l l l}
     \toprule
     \textbf{Emotion} & \multicolumn{2}{c}{\textbf{Topics}} \\ & \textbf{Highest Score} & \textbf{Lowest Score} \\
    \midrule
     Joy & 34, 94, 14, 42, 31 & 52, 99, 55, 72, 13\\  
     Anticipation & 94, 14, 31, 77, 43 & 99, 55, 72, 13, 4 \\ 
     Trust & 51, 34, 94, 7, 80 & 72, 71, 47, 13, 4\\
     Disgust & 7, 80, 95, 0, 24 & 65, 78, 48, 39, 20\\
     Suprise & 77, 43, 79, 81, 84 & 17, 10, 96, 93, 49\\
     Anger & 34, 7, 80, 95, 0 & 78, 48, 39, 20, 62\\
     Sadness & 34, 7, 80, 95, 0 & 52, 1, 82, 99, 48\\
     Fear & 7, 95, 0, 81, 92 & 17, 96, 29, 37, 35\\
     Love & 34, 7, 80, 95, 14 & 68, 36, 19, 27, 40\\
    \bottomrule
    \end{tabular}
\end{table}

At an aggregated level, the data indicate a strong dominance of Anticipation (48\%) and Joy (46\%) across all topics. The remaining 6\% of topics were primarily characterised by negative emotions, i.e. Disgust (3\%), Fear (2\%), and Anger at a minimal level of 1\%.

\begin{figure}
    \centering
    \includegraphics[width=0.99\columnwidth]{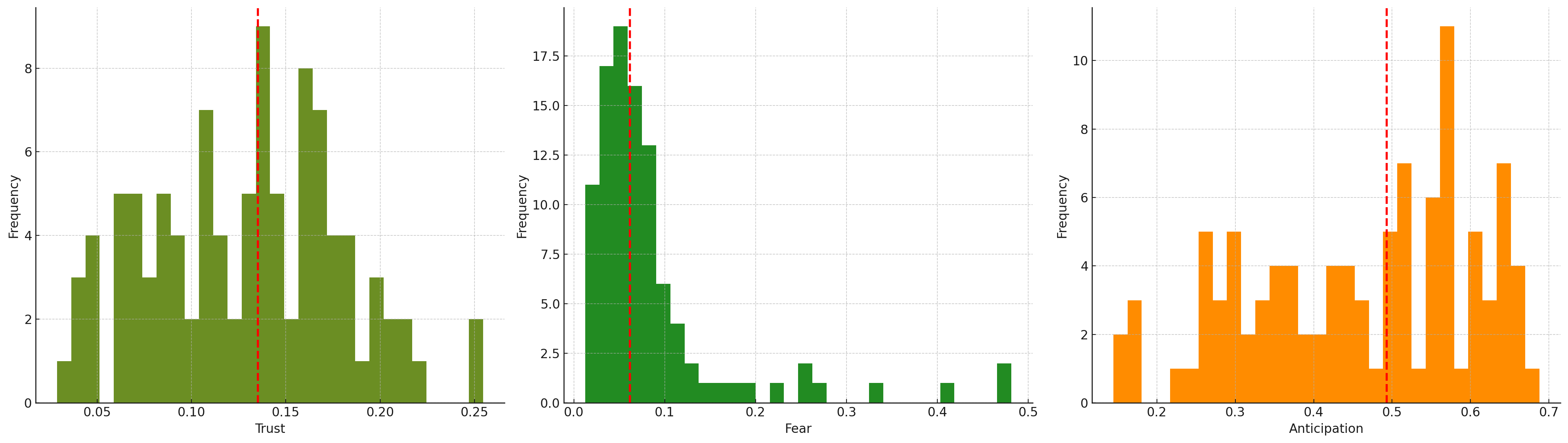}
    \caption{Distribution of Anticipation, Trust, and Fear in Topics}
    \label{fig:distribution_of_joy_trust_fear}
\end{figure}

Anticipation correlates with both Trust (leading to \textit{Hope}) and Fear (leading to Anxiety), for the frequency of these emotion scores across topics refer to \Cref{fig:distribution_of_joy_trust_fear}. The median \textit{Hope} score (0.6345) is approximately 10.33\% higher than the median Anxiety score (0.5751). The highest \textit{Hope} scores were found in topics such as "Leadership of Change and Management" (Topic 51), "CIO, Business, and Future of Work" (Topic 94), "Customer Experience and Employees" (Topic 80), "Manufacturing, Agtech, and Sustainability Innovation" (Topic 14), and "Leadership and Future of Work" (Topic 95). Topics with the lowest \textit{Hope} scores included "People Skills and Collaboration" (Topic 7), "Climate Change" (Topic 0), "Education, Economic, and Future Economy" (Topic 57), "Remote Work and Future of Work" (Topic 9), and "COVID-19 and Vaccination" (Topic 60). The highest Anxiety scores were observed in topics such as "CIO, Business, and Future of Work" (Topic 94), "Manufacturing, Agtech, and Sustainability Innovation" (Topic 14), "Customer Experience and Engagement" (Topic 31), "Quantum Computing and Latest Tech" (Topic 43), and "Innovation and Engineering in Technology" (Topic 79). Topics with the lowest Anxiety scores included "Leadership of Change and Management" (Topic 51), "People Skills and Collaboration" (Topic 7), "Customer Experience and Employees" (Topic 80), "Leadership and Future of Work" (Topic 95), and "Climate Change" (Topic 0). The non-parametric one-sided Wilcoxon signed-rank test shows the \textit{Hope} dominates over Anxiety across topics ($p = 5.58 \times 10^{-7}$).

\begin{figure}
    \centering
    \includesvg[width=0.99\columnwidth,inkscapelatex=false]{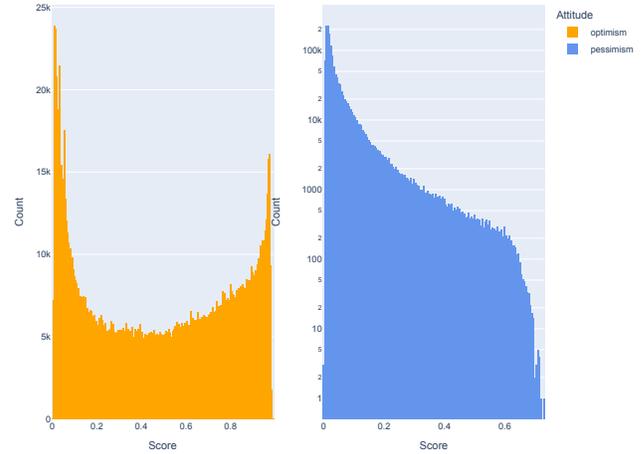}
    \caption{Distribution of Optimism and Pessimism Across the Corpus}
    \label{fig:density_optimism_pessism}
\end{figure}

\paragraph{Attitudes.} At level of Topics, the median \textit{Optimism} score (0.9979) is approximately 71.15\% higher than the median "Pessimism" score (0.5830) 
The highest \textit{Optimism} scores were found in topics such as "ReactJS, RStats, Linux, JavaScript" (Topic 35), "Quantum Computing and Tech Innovations" (Topic 43), "Leadership and Change Management" (Topic 73), "Deep Learning and Data Science" (Topic 53), and "Electric Vehicles and Auto Industry" (Topic 65). Conversely, topics with the lowest \textit{Optimism} scores include "Climate Change" (Topic 0), "Women's History Month and Time Reflections" (Topic 8), "Insurance and Manufacturing" (Topic 10), and "Year, Time, Mypov" (Topic 4). A bimodal distribution of \textit{Optimism} is observed (refer to \Cref{fig:density_optimism_pessism}). Non-tech topics like "Leadership of Change and Management" (Topic 51) and "CIO and Agile Work" (Topic 94) show the highest \textit{Optimism}, while tech topics such as "Innovation and Engineering" (Topic 79) and "Metaverse and Tech Innovation" (Topic 81) have a narrower range of \textit{Optimism} scores. 

The highest "Pessimism" scores were observed in topics such as "Artificial Intelligence and Big Data" (Topic 20), "IoT and Security Risks" (Topic 54), "Metaverse and Tech Innovation" (Topic 81), "War in Ukraine and Cybersecurity" (Topic 24), and "Lifestyle, Fashion, Art Design, Eco" (Topic 97), while the lowest scores were seen in "Women's History Month and Time Reflections" (Topic 8), "Future of Leaders and Strategy" (Topic 23), "Startups and Cryptocurrency" (Topic 41), "Politics and Policy" (Topic 39), and "Leadership and Future of Work" (Topic 95).

At the level of clusters, the median \textit{Optimism} score (0.4963) is approximately 73.29\% higher than the median "Pessimism" score (0.2864). Clusters demonstrating the highest \textit{Optimism} scores are "AI and Autonomous Robots" (Cluster 7), "AI, ML, NLP" (Cluster 1), and "AI and Healthcare" (Cluster 5), while the highest Pessimism scores are in Clusters "AI, ML, NLP" (Cluster 1), "AI and Autonomous Robots" (Cluster 7), and "AI and Healthcare" (Cluster 5).

\subsection{Validation}
We manually annotated 500 technology-related posts, selected using stratified sampling, to evaluate model performance. Posts were categorized by sentiment (positive, neutral, negative) and emotions (e.g., \textit{Hope}, Fear, Trust). Performance evaluation used Precision, Recall, and F1-Score (\Cref{tab:models_perform}).

Precision (0.87) indicates prediction accuracy - 87\% of positive sentiment predictions were correct. Recall (0.85) shows detection completeness - 85\% of true positive cases identified. F1-Score balances these metrics for overall effectiveness. The inter-annotator agreement reached a Cohen's Kappa of 0.82, demonstrating substantial reliability. These strong results validate our model's effectiveness in handling technology-specific discourse, addressing common concerns about domain adaptation in sentiment analysis.

\begin{table}[t]
\caption{Sentiment`s and Emotion`s Model Performance}
\label{tab:models_perform}
\begin{tabular}{p{2cm} p{2.5cm} p{2.5cm}}
\toprule
\textbf{Metric}         & \textbf{Sentiment Analysis} & \textbf{Emotion \newline Detection} \\ \midrule
Precision               & 0.87                       & 0.84                       \\
Recall                  & 0.85                       & 0.81                       \\
F1-Score                & 0.86                       & 0.83                       \\ \bottomrule
\end{tabular}
\end{table}

\section{Discussion}

\subsection{Anticipating Futures} 
Our analysis shows that anticipatory discourses by KOLs align closely with the concept of present future, as they often emphasize visions of transformative technological advancements. Simultaneously, the influence of contemporary events suggests that these visions actively shape the \textit{future present}, steering current debates. Among tech-driven discussions such as "Disruptv, Vimeo, and Episode Video" (Topic 25), "Climate Change" (Topic 0), "War, Ukraine, and Trump People" (Topic 83), "Machine Learning, Data Science, and Deep Learning" (Topic 6), and "IoT Community and IoT Channel" (Topic 46) generate the most dynamic engagement of KOLs, indicating their prominence in anticipatory discourse.  Conversely, topics like "Robotics, Robot, and Ronald Van Loon" (Topic 70), "COVID-19, quarantine, things to do" (Topic 99), "Healthcare, AI" (Topic  68), "Innovation engineering, energy technology" (Topic 79) and "Manufacturing, Agtech, sustainability, innovation" (Topic 14) are discussed less frequently, varying levels of societal focus. KOLs predominantly project and introduce \textit{present futures} in their posts by highlighting emerging technologies like AI, IoT, and Robotics as transformative solutions to societal challenges (e.g., climate change). These innovations are framed as key elements of a desirable and promising future \citep{adam_futures_2011}. In contrast, \textit{future presents} emerge in ongoing technological debates, such as those surrounding cybersecurity or the ethics of AI, which are shaped by the imagined long-term consequences of these technologies. These discussions represent open-ended \textit{future presents}, reflecting the interplay between immediate concerns and anticipated futures \citep{luhmann_differentiation_1982}. 

Events like the WEF and COP 27 catalyse spikes in engagement for specific topics, demonstrating how external milestones transform \textit{future presents} into focal points of public attention, action, and decision-making \citep{luhmann_differentiation_1982}. Temporal analysis demonstrates that such events influence anticipatory discourse by aligning imagined futures with immediate societal priorities. For example, during the WEF, discussions on "Climate Change" (Topic 0) reflect \textit{Optimism} about technological solutions, reinforcing \textit{present futures} as drivers of \textit{Hope} and innovation. In contrast, peaks in "War, Ukraine, and Trump People" (Topic 83) during COP 27 highlight heightened \textit{Anxiety} about geopolitical instability, illustrating how \textit{future presents} shaped by uncertainty resonate within KOL narratives. These temporal alignments underscore the dynamic role of global events in amplifying anticipatory discourse among tech influencers. By connecting societal challenges, such as climate change and geopolitical crises, with emerging technologies, KOLs mediate the intersection of \textit{Optimism} and \textit{Anxiety}, driving narratives that reflect the urgency and complexity of addressing future-oriented challenges \citep{allcott_social_2017, falkenberg_growing_2022, hopke_visualizing_2018, liang_dynamics_2023}. This analysis highlights how external milestones not only structure the rhythm of anticipatory discourse but also reinforce the duality of KOLs’ roles as narrators of imagined futures and interpreters of present realities.

Inter-topic dynamics illustrate the interconnectedness of themes within the discourse, reflecting how different technological advancements are often discussed in relation to one another. This mapping provides valuable insights into the structural organization of the analysed discussions, highlighting areas of thematic overlap and shared relevance \citep{porter_tech_2005}. For example, AI frequently intersects with topics like Big Data, automation, and IoT, demonstrating its central role in shaping multifaceted conversations across diverse technological domains.

\subsection{Anticipating Trends: Influencers and Technologies}
KOLs like Jamesvgingerich, digital\_trans4m, and kirkdborne dominate discourse in fields such as "Climate Change" (Topic 0) and "Robotics, Robot, and Engineering" (Topic 88). By fostering communities and gaining recognition in distinct subjects, they significantly contribute to shaping anticipatory discourse \citep{furini_x_2024, haupt_facebook_2021, lichti_decentralized_2023}. These influencers establish an influential `anticipatory news infrastructure` \citep{ananny_anticipatory_2020}, by curating content that integrates anticipatory emotions into narratives of potential futures.  Their influence extends beyond commentary, catalysing interest and confidence in the transformative potential of new technologies \citep{valente_identifying_2007}. Tech influencers, through their anticipatory discourse and emotional framing, actively shape future narratives. By highlighting specific technological advancements, they position themselves as future creators, driving fostering debate and promoting acceptance of emerging technologies.

The concentration of discourse power among a few influencers underscores their central role in content creation and dissemination \citep{martin_getting_2022, oueslati_recognition_2023}. Their high contribution to the number of posts highlights their disproportionate influence on framing discussions, which raises questions about the diversity and inclusivity of the tech discourse. This imbalance suggests that a few dominant voices significantly shape the anticipatory visions of technological futures within the dataset.

\subsection{Anticipatory Emotions, Sentiments and Attitudes}
A high presence of Anticipation as an emotion (48\%) exceeds the English post average (13.9\%) \citep{mohammad_semeval-2018_2018} and strongly suggests that the dataset is well-suited for the analysis. Anticipation, often linked to future positive events \citep{tenhouten_emotions_2023}, and Joy, as a response to favourable stimuli \citep{emmons_joy_2020}, dominate the corpus. Conversely, Fear, Disgust, and Anger frequently coexist in discussions on moral or ethical issues, such as war, politics, and data privacy \citep{cannon_james-lange_1927, zhan_distinctive_2015, russell_bodily_2013}. These emotions are particularly evident in debates about cybersecurity, where Anger and Fear are triggered by concerns over privacy and threats to security. Such emotions play a critical role in shaping public perception, influencing decision-making and behaviour \citep{butz_anticipatory_2007, hoffmann_anticipatory_2003}.

We demonstrated the dominance of anticipatory emotions \citep{castelfranchi_anticipation_2011, plutchik_nature_2001, feil_anticipatory_2022} by highlighting the prevalence of uncertainty-related emotions, such as \textit{Hope} and \textit{Anxiety} \citep{gordon_emotions_1969, gordon_structure_1987}, in the posts published by KOLs. These emotions prominently reflect their priorities and concerns regarding technological advancements and uncertainties. \textit{Hope} prevails over \textit{Anxiety} throughout the corpus. Topics with the highest \textit{Hope} scores include "Leadership of Change and Management" (Topic 51), "CIO, Business, Future of Work" (Topic 94), "Customer Experience and Employees" (Topic 80), "Manufacturing, Agtech, and Sustainability Innovation" (Topic 14), "Leadership, Leaders, Future of Work" (Topic 95). Trust fosters \textit{Hope} \citep{pleeging_characterizing_2022, pleeging_relations_2021}. \textit{Hope} is inherently tied to anticipation, fostering a forward-looking perspective that helps individuals manage present challenges by focusing on potential future successes \citep{tenhouten_emotions_2023}. This combination can drive proactive behaviour and resilience in the face of adversity \citep{pleeging_relations_2021}. The emphasis on \textit{Hope} in anticipatory discourses underscores the dominant role of imagined futures in shaping narratives, aligning with the concept of "present future". 

The highest \textit{Anxiety} scores include topics such as "CIO, Business, Future of Work" (Topic 94), "Manufacturing, Agtech, and Sustainability Innovation" (Topic 14), "Customer Experience and Engagement" (Topic 31), Quantum Computing and Latest Tech" (Topic 43), and "Innovation and Engineering in Technology" (Topic 79).In contrast, topics with the lowest \textit{Anxiety} scores encompass "Leadership of Change and Management" (Topic 51), "People Skills and Collaboration" (Topic 7), "Customer Experience and Employees" (Topic 80), "Leadership, Leaders, Future of Work" (Topic 95), and "Climate Change" (Topic 0). The expectations related to these topics, as set by KOLs, introduce Fear, which is a response triggered by the appraisal of imminent danger \citep{kurth_moral_2015, kurth_anxious_2018}. This Fear can subsequently activate preparatory behaviours for an anticipated future, as observed in \citep{castelfranchi_anticipation_2011}.

While positive emotions still dominate, most posts exhibit a neutral sentiment. This apparent discrepancy can be attributed to the distinct dimensions captured by sentiment and emotion analysis. Neutral sentiment reflects the linguistic tone of the posts, where influencers often convey information with minimal positive or negative polarity. This is particularly common in professional contexts, such as discussions of technology, where the goal is to inform or describe objectively rather than to express overt opinions. Despite this neutrality, the prevalence of the KOLs' posts reflects their \textit{Optimism} about technological advancements, aligning with the observations of \citet{tenhouten_emotions_2023}, and a trend towards techno-optimism \citep{konigs_what_2022}. Simultaneously, their posts also express \textit{Pessimism} regarding geopolitical uncertainty, as suggested by \citet{kurth_anxious_2018}. This connection confirms the role of anticipation in shaping decision-making and public discourse \citep{moore_data-frame_2011, tavory_coordinating_2013}.

Fear, Disgust, and Anger often coexist, indicating that discussions touch on moral or ethical issues \citep{cannon_james-lange_1927, zhan_distinctive_2015, russell_bodily_2013}. For example, war, politics, and data privacy discussions trigger Disgust and Anger, while "Cybersecurity" (e.g. Topic 2) debates elicit Anger and Fear. These emotions, especially in morally charged situations, have distinct triggers and outcomes and are known to predict moral outrage \citep{hutcherson_moral_2011, salerno_interactive_2013}. Moreover, they play a key role in shaping decision-making and behaviour \citep{butz_anticipatory_2007, butz_behavioral_2017, feil_anticipatory_2022, hoffmann_anticipatory_2003, odou_how_2020}, influencing choices and actions based on future expectations, leading to optimism and motivation or caution.


\section{Limitations and Future Work}

\subsection{Large-Scale Discourse Analysis}  
This study employs a large-scale, quantitative approach to analysing anticipatory discourse, utilizing advanced text mining techniques such as BERTopic modeling and emotion detection. While this method is well-suited for identifying macro-level patterns and trends, it does not provide the granular depth associated with qualitative analysis, such as examining individual posts or contextual nuances. Future research could address these limitations by employing qualitative methods to explore individual posts or specific themes in greater depth.

\subsection{Access Limitations Due to X's Policy} Due to restrictions in X's API during data collection, this study could not incorporate direct behavioural engagement metrics, such as likes, shares, and comments. While post volume served as a practical alternative to capture user activity, it does not fully capture the multifaceted nature of influence. Incorporating replies and comments could offer a more nuanced view of public engagement with technological discourse. Additionally, integrating them alongside topic and emotion analysis could deepen our understanding of audience engagement and influence, providing a more comprehensive perspective on the dynamics of influence and impact.

\subsection{Domain Gaps in Tech Discourse} The nuanced nature of technology discourse presents challenges for NLP models, even when trained on social media data. Future research could address these challenges by fine-tuning models with additional training data from technology-focused posts. This approach would help to bridge the domain mismatch and improve model performance in identifying anticipatory discussions in technological contexts. Additionally, the development of custom lexicons or embeddings tailored to the unique vocabulary and structure of anticipatory discourse in technology could further enhance analytical accuracy and reliability.

\subsection{Longitudinal and Comparative Analyses} Expanding this research to include longitudinal and comparative analyses could offer valuable insights into how technology-related anticipatory discourse evolves over time and across different social media platforms. Such studies could explore the interplay between user characteristics, engagement trends, and emergent topics, providing a deeper understanding of the factors driving public perceptions of technological future.

\subsection{Policy Implications} 

While this study does not directly inform policy, future research could examine how these discourse patterns guide policymakers in fostering inclusivity, addressing ethical challenges, and balancing technological optimism with critical societal reflection.

\section{Conclusion}
This study examined anticipatory discourse on tech-driven futures by analysing 1.5 million posts from 400 KOLs published on the X platform from to 2021 and 2023. Advanced text mining techniques were employed, including BERTopic modelling, as well as sentiment, emotion, and attitude analyses. It allowed us to indicate 100 distinct technological futures (topics) introduced by KOLs to a public audience. Our findings highlight the dual role of KOLs in shaping anticipatory discourse, reflecting the interplay between \textit{present futures}—optimistic projections of transformative technologies like AI and IoT—, where these visions influence current debates on geopolitical and societal challenges. Topics such as "War, Ukraine, and Trump People" and "COVID-19, quarantine, things to do" were marked by heightened \textit{Anxiety}, while others, like "Machine Learning, Data Science, and Deep Learning," highlighted \textit{Optimism} about technological innovation. By framing technological advancements as solutions to societal challenges while downplaying concerns like privacy or job displacement, KOLs act as mediators of societal narratives. This study positions them as bridges between imagined futures and current realities, balancing tangible predictions with speculative foresight. These findings underscore the critical role of KOLs in steering societal attention, shaping public sentiment, and driving discourse on emerging technologies.

This research emphasizes at least two substantial theoretical and practical implications. Theoretically, the analysis of emotion correlations suggests that Anticipation enhances positive feelings (like Trust) while also potentially introducing mild concerns (like Fear). This indicates that Anticipation and Trust together can reliably measure \textit{Hope}. Secondly, the relationship between Anticipation and Fear although weaker, can still be used to measure \textit{Anxiety}. The moderate negative correlation between Anticipation and Sadness supports using them together to measure \textit{Pessimism}, as increased Anticipation is associated with decreased Sadness. Finally, given the very weak correlation between Anticipation and `Joy,` this combination may not be as effective for measuring \textit{Optimism}. It is worth considering other combinations or additional factors to measure \textit{Optimism} more accurately. 

Future research should expand upon these findings by incorporating qualitative methods and exploring cross-platform comparisons to further unravel the complexities of anticipatory technological narratives and their implications.

\section*{Acknowledgements}
This work received financial support from 
the Faculty of Information Technology of the Czech Technical University in Prague
and the SWPS University.\\

\bibliographystyle{unsrtnat}


\begin{thebibliography}{88}
\providecommand{\natexlab}[1]{#1}
\providecommand{\url}[1]{\texttt{#1}}
\expandafter\ifx\csname urlstyle\endcsname\relax
  \providecommand{\doi}[1]{doi: #1}\else
  \providecommand{\doi}{doi: \begingroup \urlstyle{rm}\Url}\fi

\bibitem[Gordon(1969)]{gordon_emotions_1969}
Robert~M Gordon.
\newblock Emotions and knowledge.
\newblock \emph{The Journal of Philosophy}, 66\penalty0 (13):\penalty0 408--413, 1969.
\newblock ISSN 0022-362X.
\newblock Publisher: JSTOR.

\bibitem[Gordon(1987)]{gordon_structure_1987}
Robert~M Gordon.
\newblock \emph{The structure of emotions: {Investigations} in cognitive philosophy}.
\newblock Cambridge University Press, 1987.
\newblock ISBN 0-521-39568-2.

\bibitem[Castelfranchi and Miceli(2011)]{castelfranchi_anticipation_2011}
Cristiano Castelfranchi and Maria Miceli.
\newblock Anticipation and {Emotion}.
\newblock In Roddy Cowie, Catherine Pelachaud, and Paolo Petta, editors, \emph{Emotion-{Oriented} {Systems}: {The} {Humaine} {Handbook}}, pages 483--500. Springer Berlin Heidelberg, Berlin, Heidelberg, 2011.
\newblock ISBN 978-3-642-15184-2.
\newblock \doi{10.1007/978-3-642-15184-2_25}.
\newblock URL \url{https://doi.org/10.1007/978-3-642-15184-2_25}.

\bibitem[Feil et~al.(2022)Feil, Weyland, Fritsch, Wäsche, and Jekauc]{feil_anticipatory_2022}
Katharina Feil, Susanne Weyland, Julian Fritsch, Hagen Wäsche, and Darko Jekauc.
\newblock Anticipatory and {Anticipated} {Emotions} in {Regular} and {Non}-regular {Exercisers} – {A} {Qualitative} {Study}.
\newblock \emph{Frontiers in Psychology}, 13, 2022.
\newblock ISSN 1664-1078.
\newblock \doi{10.3389/fpsyg.2022.929380}.
\newblock URL \url{https://www.frontiersin.org/articles/10.3389/fpsyg.2022.929380}.

\bibitem[Moore and Hoffman(2011)]{moore_data-frame_2011}
David~T. Moore and Robert~R. Hoffman.
\newblock Data-{Frame} {Theory} of {Sensemaking} as a {Best} {Model} for {Intelligence}.
\newblock \emph{American Intelligence Journal}, 29\penalty0 (2):\penalty0 145--158, 2011.
\newblock ISSN 0883072X.
\newblock URL \url{http://www.jstor.org/stable/26201963}.
\newblock Publisher: National Military Intelligence Association.

\bibitem[Grabenhorst et~al.(2021)Grabenhorst, Maloney, Poeppel, and Michalareas]{grabenhorst_two_2021}
Matthias Grabenhorst, Laurence~T. Maloney, David Poeppel, and Georgios Michalareas.
\newblock Two sources of uncertainty independently modulate temporal expectancy.
\newblock \emph{Proceedings of the National Academy of Sciences}, 118\penalty0 (16):\penalty0 e2019342118, April 2021.
\newblock \doi{10.1073/pnas.2019342118}.
\newblock URL \url{https://doi.org/10.1073/pnas.2019342118}.
\newblock Publisher: Proceedings of the National Academy of Sciences.

\bibitem[Hirsh et~al.(2012)Hirsh, Mar, and Peterson]{hirsh_psychological_2012}
Jacob~B. Hirsh, Raymond~A. Mar, and Jordan~B. Peterson.
\newblock Psychological entropy: {A} framework for understanding uncertainty-related anxiety.
\newblock \emph{Psychological Review}, 119\penalty0 (2):\penalty0 304--320, 2012.
\newblock ISSN 1939-1471(Electronic),0033-295X(Print).
\newblock \doi{10.1037/a0026767}.
\newblock Place: US Publisher: American Psychological Association.

\bibitem[FeldmanHall and Shenhav(2019)]{feldmanhall_resolving_2019}
Oriel FeldmanHall and Amitai Shenhav.
\newblock Resolving uncertainty in a social world.
\newblock \emph{Nat. Hum. Behav.}, 3\penalty0 (5):\penalty0 426--435, April 2019.
\newblock ISSN 2397-3374.
\newblock \doi{10.1038/s41562-019-0590-x}.
\newblock URL \url{http://dx.doi.org/10.1038/s41562-019-0590-x}.
\newblock Publisher: Springer Science and Business Media LLC.

\bibitem[Koivunen et~al.(2023)Koivunen, Nikunen, Hokkanen, Jaaksi, Lehtinen, Soronen, Talvitie-Lamberg, and Valtonen]{koivunen_anticipation_2023}
Anu Koivunen, Kaarina Nikunen, Julius Hokkanen, Vilja Jaaksi, Vilma Lehtinen, Anne Soronen, Karoliina Talvitie-Lamberg, and Sanna Valtonen.
\newblock Anticipation as {Platform} {Power}: {The} {Temporal} {Structuring} of {Digital} {Everyday} {Life}.
\newblock \emph{Television \& New Media}, page 15274764231178228, June 2023.
\newblock ISSN 1527-4764.
\newblock \doi{10.1177/15274764231178228}.
\newblock URL \url{https://doi.org/10.1177/15274764231178228}.\newblock Publisher: SAGE Publications.

\bibitem[Tavory and Eliasoph(2013)]{tavory_coordinating_2013}
Iddo Tavory and Nina Eliasoph.
\newblock Coordinating {Futures}: {Toward} a {Theory} of {Anticipation}.
\newblock \emph{American Journal of Sociology}, 118\penalty0 (4):\penalty0 908--942, 2013.
\newblock ISSN 00029602, 15375390.
\newblock \doi{10.1086/668646}.
\newblock URL \url{http://www.jstor.org/stable/10.1086/668646}.
\newblock Publisher: The University of Chicago Press.

\bibitem[Acerbi(2022)]{acerbi_storytelling_2022}
Alberto Acerbi.
\newblock From {Storytelling} to {Facebook}.
\newblock \emph{Human Nature}, 33:\penalty0 132--144, 2022.
\newblock ISSN 1936-4776.
\newblock \doi{10.1007/s12110-022-09423-1}.
\newblock URL \url{https://doi.org/10.1007/s12110-022-09423-1}.

\bibitem[Acerbi(2019)]{acerbi_cultural_2019}
Alberto Acerbi.
\newblock \emph{Cultural {Evolution} in the {Digital} {Age}}.
\newblock Oxford University Press, Oxford, 2019.

\bibitem[Jin and Phua(2014)]{jin_following_2014}
Seung-A~Annie Jin and Joe Phua.
\newblock Following {Celebrities}’ {posts} {About} {Brands}: {The} {Impact} of {X}-{Based} {Electronic} {Word}-of-{Mouth} on {Consumers}’ {Source} {Credibility} {Perception}, {Buying} {Intention}, and {Social} {Identification} {With} {Celebrities}.
\newblock \emph{Journal of Advertising}, 43\penalty0 (2):\penalty0 181--195, April 2014.
\newblock ISSN 0091-3367.
\newblock \doi{10.1080/00913367.2013.827606}.
\newblock URL \url{https://doi.org/10.1080/00913367.2013.827606}.
\newblock Publisher: Routledge.

\bibitem[Landowska et~al.(2023)Landowska, Robak, and Skorski]{landowska_what_2023}
Alina Landowska, Marek Robak, and Maciej Skorski.
\newblock What {Twitter} data tell us about the future?
\newblock 2023.
\newblock \doi{10.48550/ARXIV.2308.02035}.
\newblock URL \url{http://dx.doi.org/10.48550/ARXIV.2308.02035}.
\newblock Publisher: arXiv.

\bibitem[Spry et~al.(2011)Spry, Pappu, and Bettina~Cornwell]{spry_celebrity_2011}
Amanda Spry, Ravi Pappu, and T.~Bettina~Cornwell.
\newblock Celebrity endorsement, brand credibility and brand equity.
\newblock \emph{European Journal of Marketing}, 45\penalty0 (6):\penalty0 882--909, January 2011.
\newblock ISSN 0309-0566.
\newblock \doi{10.1108/03090561111119958}.
\newblock URL \url{https://doi.org/10.1108/03090561111119958}.
\newblock Publisher: Emerald Group Publishing Limited.

\bibitem[Tur et~al.(2022)Tur, Harstad, and Antonakis]{tur_effect_2022}
Benjamin Tur, Jennifer Harstad, and John Antonakis.
\newblock Effect of charismatic signaling in social media settings: {Evidence} from {TED} and {Twitter}.
\newblock \emph{The Leadership Quarterly}, 33\penalty0 (5):\penalty0 101476, October 2022.
\newblock ISSN 1048-9843.
\newblock \doi{10.1016/j.leaqua.2020.101476}.
\newblock URL \url{https://www.sciencedirect.com/science/article/pii/S104898432030103X}.

\bibitem[Sadiq and Khan(2018)]{sadiq_self-driving_2018}
Rizwan Sadiq and Mohsin Khan.
\newblock Analyzing self-driving cars on {Twitter}.
\newblock \emph{CoRR}, abs/1804.04058, 2018.
\newblock \url{http://arxiv.org/abs/1804.04058}.

\bibitem[Miyazaki et~al.(2023)Miyazaki, Murayama, Uchiba, An, and Kwak]{miyazaki_public_2023}
Kunihiro Miyazaki, Taichi Murayama, Takayuki Uchiba, Jisun An, and Haewoon Kwak.
\newblock Public perception of generative {AI} on {Twitter}: {An} empirical study based on occupation and usage.
\newblock \emph{arXiv preprint}, 2023.
\newblock URL \url{https://arxiv.org/abs/2305.09537}.


\bibitem[Adam(1990)]{adam_time_1990}
Barbara Adam.
\newblock \emph{Time and {Social} {Theory}}.
\newblock Polity, Cambridge, 1990.

\bibitem[Bergmann(1992)]{bergmann_problem_1992}
Werner Bergmann.
\newblock The {Problem} of {Time} in {Sociology}.
\newblock \emph{Time \& Society}, 1\penalty0 (1):\penalty0 81--134, January 1992.
\newblock ISSN 0961-463X.
\newblock \doi{10.1177/0961463x92001001007}.
\newblock URL \url{http://dx.doi.org/10.1177/0961463x92001001007}.
\newblock Publisher: SAGE Publications.

\bibitem[Emirbayer and Mische(1998)]{emirbayer_what_1998}
Mustafa Emirbayer and Ann Mische.
\newblock What {Is} {Agency}?
\newblock \emph{American Journal of Sociology}, 103\penalty0 (4):\penalty0 962--1023, January 1998.
\newblock ISSN 0002-9602.
\newblock \doi{10.1086/231294}.
\newblock URL \url{https://doi.org/10.1086/231294}.
\newblock Publisher: The University of Chicago Press.

\bibitem[Abbott(2001)]{abbott_time_2001}
Andrew Abbott.
\newblock \emph{Time {Matters}: {On} {Theory} and {Method}}.
\newblock University of Chicago Press, Chicago, 2001.

\bibitem[Mische(2009)]{mische_projects_2009}
Ann Mische.
\newblock Projects and {Possibilities}: {Researching} {Futures} in {Action}.
\newblock \emph{Sociological Forum}, 24\penalty0 (3):\penalty0 694--704, 2009.
\newblock ISSN 08848971, 15737861.
\newblock \doi{10.1111/j.1573-7861.2009.01127.x}.
\newblock Publisher: [Wiley, Springer].

\bibitem[Egbert and Baker(2019)]{egbert_using_2019}
Jesse Egbert and Paul Baker, editors.
\newblock \emph{Using corpus methods to triangulate linguistic analysis}.
\newblock Routledge, September 2019.
\newblock ISBN 978-1-315-11246-6.
\newblock \doi{10.4324/9781315112466}.
\newblock URL \url{http://dx.doi.org/10.4324/9781315112466}.

\bibitem[Luhmann(1982)]{luhmann_differentiation_1982}
Niklas Luhmann.
\newblock \emph{The differentiation of society}.
\newblock Columbia University Press, New York, 1982.

\bibitem[Adam and Groves(2011)]{adam_futures_2011}
Barbara Adam and Chris Groves.
\newblock Futures {Tended}: {Care} and {Future}-{Oriented} {Responsibility}.
\newblock \emph{Bulletin of Science, Technology \& Society}, 31\penalty0 (1):\penalty0 17--27, February 2011.
\newblock ISSN 0270-4676.
\newblock \doi{10.1177/0270467610391237}.
\newblock URL \url{https://doi.org/10.1177/0270467610391237}.
\newblock Publisher: SAGE Publications Inc.

\bibitem[Poli(2017)]{poli_introduction_2017}
Roberto Poli.
\newblock \emph{Introduction to {Anticipation} {Studies}}, volume~1.
\newblock Springer International Publishing AG, Electronic, January 2017.
\newblock ISBN 978-3-319-63021-2.
\newblock \doi{10.1007/978-3-319-63023-6}.

\bibitem[Poli(2014)]{poli_anticipation_2014}
Roberto Poli.
\newblock Anticipation: {A} {New} {Thread} for the {Human} and {Social} {Sciences}?
\newblock \emph{PromotingCADMUS: Leadership in Thought that Leads to Action}, 2\penalty0 (3):\penalty0 23--36, 2014.

\bibitem[Adam and Groves(2007)]{adam_future_2007}
Barbara Adam and Chris Groves, editors.
\newblock \emph{Future matters. {Action}, {Knowledge}, {Ethic}}.
\newblock Brill, Leiden \& Boston, 2007.
\newblock ISBN 90-04-16177-5.

\bibitem[Miller(2007)]{miller_futures_2007}
Riel Miller.
\newblock Futures {Literacy}: {A} {Hybrid} {Strategic} {Scenario} {Method}.
\newblock \emph{Futures}, 39\penalty0 (4):\penalty0 341--362, 2007.

\bibitem[Beckert(2013)]{beckert_capitalism_2013}
Jens Beckert.
\newblock Capitalism as a {System} of {Expectations}: {Toward} a {Sociological} {Microfoundation} of {Political} {Economy}.
\newblock \emph{Politics \& Society}, 41\penalty0 (3):\penalty0 323--350, September 2013.
\newblock ISSN 0032-3292.
\newblock \doi{10.1177/0032329213493750}.
\newblock URL \url{https://doi.org/10.1177/0032329213493750}.
\newblock Publisher: SAGE Publications Inc.

\bibitem[Bagozzi et~al.(1998)Bagozzi, Baumgartner, and Pieters]{bagozzi_goal-directed_1998}
Richard~P. Bagozzi, Hans~Rudolf Baumgartner, and Rik Pieters.
\newblock Goal-directed {Emotions}.
\newblock \emph{Cognition \& Emotion}, 12:\penalty0 1--26, 1998.
\newblock \doi{10.1111/acer.14513}.
\newblock URL \url{https://api.semanticscholar.org/CorpusID:143822376}.

\bibitem[Baumgartner et~al.(2008)Baumgartner, Pieters, and Bagozzi]{baumgartner_future-oriented_2008}
Hans Baumgartner, Rik Pieters, and Richard~P. Bagozzi.
\newblock Future-oriented emotions: conceptualization and behavioral effects.
\newblock \emph{European Journal of Social Psychology}, 38\penalty0 (4):\penalty0 685--696, June 2008.
\newblock ISSN 0046-2772.
\newblock \doi{10.1002/ejsp.467}.
\newblock URL \url{https://doi.org/10.1002/ejsp.467}.
\newblock Publisher: John Wiley \& Sons, Ltd.

\bibitem[Perugini and Bagozzi(2001)]{perugini_role_2001}
Marco Perugini and Richard Bagozzi.
\newblock The role of desires and anticipated emotions in goal-directed behaviors: {Broadening} and deepening the theory of planned behavior.
\newblock \emph{British Journal of Social Psychology}, 40:\penalty0 79--98, March 2001.
\newblock \doi{10.1348/014466601164704}.

\bibitem[Bruns et~al.(2015)Bruns, Enli, Skogerbø, Larsson, and Christensen]{bruns_routledge_2015}
Axel Bruns, Gunn Enli, Eli Skogerbø, Anders Olof Larsson, and Christian Christensen.
\newblock \emph{The Routledge companion to social media and politics}.
\newblock Routledge, New York, NY, 2015.
\newblock ISBN 9781315716299.
\newblock \doi{10.4324/9781315716299}.
\newblock URL \url{http://dx.doi.org/10.4324/9781315716299}.



\bibitem[MacLeod(2017)]{macleod_anticipatory_2017}
Andrew MacLeod.
\newblock Anticipatory feelings.
\newblock In Andrew MacLeod, editor, \emph{Prospection, well-being, and mental health}, page~0. Oxford University Press, February 2017.
\newblock ISBN 978-0-19-872504-6.
\newblock \doi{10.1093/med:psych/9780198725046.003.0005}.
\newblock URL \url{https://doi.org/10.1093/med:psych/9780198725046.003.0005}.

\bibitem[Vazard(2024)]{vazard_feeling_2024}
Juliette Vazard.
\newblock Feeling the {Unknown}: {Emotions} of {Uncertainty} and {Their} {Valence}.
\newblock \emph{Erkenntnis}, 89\penalty0 (4):\penalty0 1275--1294, April 2024.
\newblock ISSN 1572-8420.
\newblock \doi{10.1007/s10670-022-00583-1}.
\newblock URL \url{https://doi.org/10.1007/s10670-022-00583-1}.

\bibitem[Mowrer(1960)]{mowrer_learning_1960}
Orval~H. Mowrer.
\newblock \emph{Learning {Theory} and {Behavior}}.
\newblock Wiley, New York, 1960.

\bibitem[Plutchik(1980)]{plutchik_general_1980}
Robert Plutchik.
\newblock A general psychoevolutionary theory of emotion.
\newblock In Robert Plutchik and Henry Kellerman, editors, \emph{Emotion: {Theory}, {Research} and {Experience}. {Theories} of {Emotions}.}, volume~1, pages 3--33. Academic Press, New York, 1980.


\bibitem[Plutchik(2001)]{plutchik_nature_2001}
Robert Plutchik.
\newblock The {Nature} of {Emotions}.
\newblock \emph{American Scientist}, 89\penalty0 (4):\penalty0 344--350, July 2001.
\newblock ISSN 00030996.
\newblock URL \url{http://www.jstor.org/stable/27857503}.
\newblock 344.

\bibitem[JustAnotherArchivist(2023)]{justanotherarchivist_snscrape_2023}
JustAnotherArchivist.
\newblock snscrape, March 2023.
\newblock URL \url{https://github.com/JustAnotherArchivist/snscrape.git}.

\bibitem[Bogdanowicz and Guan(2022)]{bogdanowicz_dynamic_2022}
Alexander Bogdanowicz and ChengHe Guan.
\newblock Dynamic topic modeling of twitter data during the {COVID}-19 pandemic.
\newblock \emph{PLOS ONE}, 17\penalty0 (5):\penalty0 e0268669, May 2022.
\newblock ISSN 1932-6203.
\newblock \doi{10.1371/journal.pone.0268669}.
\newblock URL \url{https://dx.plos.org/10.1371/journal.pone.0268669}.

\bibitem[Egger and Yu(2022)]{egger_topic_2022}
Roman Egger and Joanne Yu.
\newblock A {Topic} {Modeling} {Comparison} {Between} {LDA}, {NMF}, {Top2Vec}, and {BERTopic} to {Demystify} {Twitter} {Posts}.
\newblock \emph{Frontiers in Sociology}, 7:\penalty0 886498, May 2022.
\newblock ISSN 2297-7775.
\newblock \doi{10.3389/fsoc.2022.886498}.
\newblock URL \url{https://www.ncbi.nlm.nih.gov/pmc/articles/PMC9120935/}.

\bibitem[Yang and Saffer(2021)]{yang_standing_2021}
Aimei Yang and Adam~J Saffer.
\newblock Standing out in a networked communication context: {Toward} a network contingency model of public attention.
\newblock \emph{New Media \& Society}, 23\penalty0 (10):\penalty0 2902--2925, October 2021.
\newblock ISSN 1461-4448.
\newblock \doi{10.1177/1461444820939445}.
\newblock URL \url{https://doi.org/10.1177/1461444820939445}.
\newblock Publisher: SAGE Publications.

\bibitem[Grootendorst(2022)]{grootendorst2022bertopic}
Maarten Grootendorst.
\newblock {BERTopic}: {Neural} topic modeling with a class-based {TF}-{IDF} procedure.
\newblock \emph{arXiv preprint arXiv:2203.05794}, 2022.

\bibitem[Sievert and Shirley(2014)]{sievert_ldavis_2014}
Carson Sievert and Kenneth Shirley.
\newblock {LDAvis}: {A} method for visualizing and interpreting topics.
\newblock In \emph{Proceedings of the {Workshop} on {Interactive} {Language} {Learning}, {Visualization}, and {Interfaces}}, pages 63--70, Baltimore, Maryland, USA, 2014. Association for Computational Linguistics.
\newblock \doi{10.3115/v1/W14-3110}.
\newblock URL \url{http://aclweb.org/anthology/W14-3110}.

\bibitem[Mabey(2021)]{mabey_pyldavis_2021}
Ben Mabey.
\newblock {pyLDAvis}, 2021.
\newblock URL \url{https://github.com/bmabey/pyLDAvis}.
\newblock original-date: 2015-04-09T22:48:03Z.

\bibitem[Casaló et~al.(2020)Casaló, Flavián, and Ibáñez-Sánchez]{casalo_influencers_2020}
Luis~V. Casaló, Carlos Flavián, and Sergio Ibáñez-Sánchez.
\newblock Influencers on {Instagram}: {Antecedents} and consequences of opinion leadership.
\newblock \emph{Journal of Business Research}, 117:\penalty0 510--519, September 2020.
\newblock ISSN 0148-2963.
\newblock \doi{10.1016/j.jbusres.2018.07.005}.
\newblock URL \url{https://www.sciencedirect.com/science/article/pii/S0148296318303187}.

\bibitem[Augustyniak et~al.(2023)Augustyniak, Woźniak, Gruza, Gramacki, Rajda, Morzy, and Kajdanowicz]{augustyniak_massively_2023}
Lukasz Augustyniak, Szymon Woźniak, Marcin Gruza, Piotr Gramacki, Krzysztof Rajda, Mikołaj Morzy, and Tomasz Kajdanowicz.
\newblock Massively {Multilingual} {Corpus} of {Sentiment} {Datasets} and {Multi}-faceted {Sentiment} {Classification} {Benchmark}.
\newblock 2023.

\bibitem[Rajda et~al.(2022)Rajda, Augustyniak, Gramacki, Gruza, Woźniak, and Kajdanowicz]{rajda_assessment_2022}
Krzysztof Rajda, Lukasz Augustyniak, Piotr Gramacki, Marcin Gruza, Szymon Woźniak, and Tomasz Kajdanowicz.
\newblock Assessment of {Massively} {Multilingual} {Sentiment} {Classiﬁers}.
\newblock In \emph{Proceedings of the 12th {Workshop} on {Computational} {Approaches} to {Subjectivity}, {Sentiment} \& {Social} {Media} {Analysis}}. Association for Computational Linguistics, 2022.
\newblock URL \url{https://aclanthology.org/2022.wassa-1.13}.

\bibitem[Mohammad et~al.(2018)Mohammad, Bravo-Marquez, Salameh, and Kiritchenko]{mohammad_semeval-2018_2018}
Saif Mohammad, Felipe Bravo-Marquez, Mohammad Salameh, and Svetlana Kiritchenko.
\newblock {SemEval}-2018 {Task} 1: {Affect} in {posts}.
\newblock In \emph{Proceedings of {The} 12th {International} {Workshop} on {Semantic} {Evaluation}}, pages 1--17, New Orleans, Louisiana, 2018. Association for Computational Linguistics.
\newblock \doi{10.18653/v1/S18-1001}.
\newblock URL \url{http://aclweb.org/anthology/S18-1001}.

\bibitem[Camacho-Collados et~al.(2022)Camacho-Collados, Rezaee, Riahi, Ushio, Loureiro, Antypas, Boisson, Espinosa~Anke, Liu, and Martínez~Cámara]{camacho-collados_postnlp_2022}
Jose Camacho-Collados, Kiamehr Rezaee, Talayeh Riahi, Asahi Ushio, Daniel Loureiro, Dimosthenis Antypas, Joanne Boisson, Luis Espinosa~Anke, Fangyu Liu, and Eugenio Martínez~Cámara.
\newblock {postNLP}: {Cutting}-{Edge} {Natural} {Language} {Processing} for {Social} {Media}.
\newblock In \emph{Proceedings of the 2022 {Conference} on {Empirical} {Methods} in {Natural} {Language} {Processing}: {System} {Demonstrations}}, pages 38--49, Abu Dhabi, UAE, 2022. Association for Computational Linguistics.
\newblock \doi{10.18653/v1/2022.emnlp-demos.5}.
\newblock URL \url{https://aclanthology.org/2022.emnlp-demos.5}.

\bibitem[Semeraro et~al.(2021)Semeraro, Vilella, and Ruffo]{semeraro_pyplutchik_2021}
Alfonso Semeraro, Salvatore Vilella, and Giancarlo Ruffo.
\newblock {PyPlutchik}: {Visualising} and comparing emotion-annotated corpora.
\newblock \emph{PLOS ONE}, 16\penalty0 (9):\penalty0 e0256503, September 2021.
\newblock ISSN 1932-6203.
\newblock \doi{10.1371/journal.pone.0256503}.
\newblock URL \url{https://dx.plos.org/10.1371/journal.pone.0256503}.

\bibitem[Bortolotti(2018)]{bortolotti_optimism_2018}
Lisa Bortolotti.
\newblock Optimism, agency, and success.
\newblock \emph{Ethical Theory and Moral Practice}, 21\penalty0 (3):\penalty0 521--535, 2018.
\newblock ISSN 1386-2820.
\newblock Publisher: Springer.

\bibitem[Jefferson et~al.(2017)Jefferson, Bortolotti, and Kuzmanovic]{jefferson_what_2017}
Anneli Jefferson, Lisa Bortolotti, and Bojana Kuzmanovic.
\newblock What is unrealistic optimism?
\newblock \emph{Consciousness and cognition}, 50:\penalty0 3--11, 2017.
\newblock ISSN 1053-8100.
\newblock Publisher: Elsevier.

\bibitem[TenHouten(2023)]{tenhouten_emotions_2023}
Warren TenHouten.
\newblock The {Emotions} of {\textit{Hope}}: {From} {Optimism} to {Sanguinity}, from {Pessimism} to {Despair}.
\newblock \emph{The American Sociologist}, 54\penalty0 (1):\penalty0 76--100, March 2023.
\newblock ISSN 1936-4784.
\newblock \doi{10.1007/s12108-022-09544-1}.
\newblock URL \url{https://doi.org/10.1007/s12108-022-09544-1}.

\bibitem[Röder et~al.(2015)Röder, Both, and Hinneburg]{roder_exploring_2015}
Michael Röder, Andreas Both, and Alexander Hinneburg.
\newblock Exploring the {Space} of {Topic} {Coherence} {Measures}.
\newblock In \emph{Proceedings of the {Eighth} {ACM} {International} {Conference} on {Web} {Search} and {Data} {Mining}}, pages 399--408, Shanghai China, February 2015. ACM.
\newblock ISBN 9781450333177.
\newblock \doi{10.1145/2684822.2685324}.
\newblock URL \url{https://dl.acm.org/doi/10.1145/2684822.2685324}.

\bibitem[Campagnolo et~al.(2022)Campagnolo, Duarte, and Dal~Bianco]{campagnolo_topic_2022}
João~Marcos Campagnolo, Denio Duarte, and Guillherme Dal~Bianco.
\newblock Topic {Coherence} {Metrics}: {How} {Sensitive} {Are} {They}?
\newblock \emph{Journal of Information and Data Management}, 13\penalty0 (4), October 2022.
\newblock ISSN 2178-7107, 2178-7107.
\newblock \doi{10.5753/jidm.2022.2181}.
\newblock URL \url{https://sol.sbc.org.br/journals/index.php/jidm/article/view/2181}.

\bibitem[Santakij et~al.(2024)Santakij, Srisuay, and Punpeng]{santakij_analyzing_2024}
Pakorn Santakij, Samai Srisuay, and Pongporn Punpeng.
\newblock Analyzing {COVID}-19 {Discourse} on {Twitter}: {Text} {Clustering} and {Classification} {Models} for {Public} {Health} {Surveillance}.
\newblock \emph{Computer Systems Science and Engineering}, 0\penalty0 (0):\penalty0 1--10, 2024.
\newblock ISSN 0267-6192.
\newblock \doi{10.32604/csse.2024.045066}.
\newblock URL \url{https://www.techscience.com/csse/online/detail/20083}.

\bibitem[Austin et~al.(2024)Austin, Makwana, Trabelsi, Largeron, and Zaïane]{austin_uncovering_2024}
Eric Austin, Shraddha Makwana, Amine Trabelsi, Christine Largeron, and Osmar~R. Zaïane.
\newblock Uncovering {Flat} and {Hierarchical} {Topics} by {Community} {Discovery} on {Word} {Co}-occurrence {Network}.
\newblock \emph{Data Science and Engineering}, 9\penalty0 (1):\penalty0 41--61, March 2024.
\newblock ISSN 2364-1541.
\newblock \doi{10.1007/s41019-023-00239-2}.
\newblock URL \url{https://doi.org/10.1007/s41019-023-00239-2}.

\bibitem[Chen et~al.(2023)Chen, Rabhi, Liao, and Al-Qudah]{chen_leveraging_2023}
Weisi Chen, Fethi Rabhi, Wenqi Liao, and Islam Al-Qudah.
\newblock Leveraging {State}-of-the-{Art} {Topic} {Modeling} for {News} {Impact} {Analysis} on {Financial} {Markets}: {A} {Comparative} {Study}.
\newblock \emph{Electronics}, 12\penalty0 (12):\penalty0 2605, January 2023.
\newblock ISSN 2079-9292.
\newblock \doi{10.3390/electronics12122605}.
\newblock URL \url{https://www.mdpi.com/2079-9292/12/12/2605}.
\newblock Number: 12 Publisher: Multidisciplinary Digital Publishing Institute.

\bibitem[Allcott and Gentzkow(2017)]{allcott_social_2017}
Hunt Allcott and Matthew Gentzkow.
\newblock Social {Media} and {Fake} {News} in the 2016 {Election}.
\newblock \emph{Journal of Economic Perspectives}, 31:\penalty0 211--236, May 2017.
\newblock \doi{10.1257/jep.31.2.211}.

\bibitem[Falkenberg et~al.(2022)Falkenberg, Galeazzi, Torricelli, Di~Marco, Larosa, Sas, Mekacher, Pearce, Zollo, Quattrociocchi, and Baronchelli]{falkenberg_growing_2022}
Max Falkenberg, Alessandro Galeazzi, Maddalena Torricelli, Niccolò Di~Marco, Francesca Larosa, Madalina Sas, Amin Mekacher, Warren Pearce, Fabiana Zollo, Walter Quattrociocchi, and Andrea Baronchelli.
\newblock Growing polarization around climate change on social media.
\newblock \emph{Nature Climate Change}, 12\penalty0 (12):\penalty0 1114--1121, December 2022.
\newblock ISSN 1758-6798.
\newblock \doi{10.1038/s41558-022-01527-x}.
\newblock URL \url{https://doi.org/10.1038/s41558-022-01527-x}.

\bibitem[Hopke and Hestres(2018)]{hopke_visualizing_2018}
Jill~E. Hopke and Luis~E. Hestres.
\newblock Visualizing the {Paris} {Climate} {Talks} on {Twitter}: {Media} and {Climate} {Stakeholder} {Visual} {Social} {Media} {During} {COP21}.
\newblock \emph{Social Media + Society}, 4\penalty0 (3):\penalty0 2056305118782687, July 2018.
\newblock ISSN 2056-3051.
\newblock \doi{10.1177/2056305118782687}.
\newblock URL \url{https://doi.org/10.1177/2056305118782687}.
\newblock Publisher: SAGE Publications Ltd.

\bibitem[Liang and Lu(2023)]{liang_dynamics_2023}
Fan Liang and Shuning Lu.
\newblock The {Dynamics} of {Event}-{Based} {Political} {Influencers} on {Twitter}: {A} {Longitudinal} {Analysis} of {Influential} {Accounts} {During} {Chinese} {Political} {Events}.
\newblock \emph{Social Media + Society}, 9\penalty0 (2):\penalty0 20563051231177946, April 2023.
\newblock ISSN 2056-3051.
\newblock \doi{10.1177/20563051231177946}.
\newblock URL \url{https://doi.org/10.1177/20563051231177946}.
\newblock Publisher: SAGE Publications Ltd.

\bibitem[Porter and Cunningham(2005)]{porter_tech_2005}
Alan~L. Porter and Scott~W. Cunningham.
\newblock \emph{Tech {Mining}: {Exploiting} {New} {Technologies} for {Competitive} {Advantage}}.
\newblock John Wiley \& Sons, Inc., Hoboken, New Jersey, January 2005.

\bibitem[Furini(2024)]{furini_x_2024}
Marco Furini.
\newblock X as a {Passive} {Sensor} to {Identify} {Opinion} {Leaders}: {A} {Novel} {Method} for {Balancing} {Visibility} and {Community} {Engagement}.
\newblock \emph{Sensors}, 24\penalty0 (2), 2024.
\newblock ISSN 1424-8220.
\newblock \doi{10.3390/s24020610}.

\bibitem[Haynes(2013)]{haynes_bonferroni_2013}
William Haynes.
\newblock Bonferroni {Correction}.
\newblock In: Werner Dubitzky, Olaf Wolkenhauer, Kwang-Hyun Cho, Hiroki Yokota (eds.), \emph{Encyclopedia of Systems Biology}. Springer, New York, NY, 2013.
\newblock \doi{10.1007/978-1-4419-9863-7_1213}.
\newblock URL \url{https://doi.org/10.1007/978-1-4419-9863-7_1213}.

\bibitem[Haupt(2021)]{haupt_facebook_2021}
Joachim Haupt.
\newblock Facebook futures: {Mark} {Zuckerberg}’s discursive construction of a better world.
\newblock \emph{New Media \& Society}, 23\penalty0 (2):\penalty0 237--257, February 2021.
\newblock ISSN 1461-4448.
\newblock \doi{10.1177/1461444820929315}.
\newblock URL \url{https://doi.org/10.1177/1461444820929315}.
\newblock Publisher: SAGE Publications.

\bibitem[Lichti et~al.(2023)Lichti, Ademi, and Tumasjan]{lichti_decentralized_2023}
Constantin Lichti, Endrit Ademi, and Andranik Tumasjan.
\newblock \emph{Decentralized {Opinion} {Leadership}: {A} {Study} of {Crypto} {Influencers} in the {Twitter} {Discourse} on {Bitcoin}}.
\newblock September 2023.

\bibitem[Ananny and Finn(2020)]{ananny_anticipatory_2020}
Mike Ananny and Megan Finn.
\newblock Anticipatory news infrastructures: {Seeing} journalism’s expectations of future publics in its sociotechnical systems.
\newblock \emph{New Media \& Society}, 22\penalty0 (9):\penalty0 1600--1618, September 2020.
\newblock ISSN 1461-4448.
\newblock \doi{10.1177/1461444820914873}.
\newblock URL \url{https://doi.org/10.1177/1461444820914873}.
\newblock Publisher: SAGE Publications.

\bibitem[Valente and Pumpuang(2007)]{valente_identifying_2007}
Thomas~W. Valente and Patchareeya Pumpuang.
\newblock Identifying {Opinion} {Leaders} to {Promote} {Behavior} {Change}.
\newblock \emph{Health Education \& Behavior}, 34\penalty0 (6):\penalty0 881--896, December 2007.
\newblock ISSN 1090-1981.
\newblock \doi{10.1177/1090198106297855}.
\newblock URL \url{https://doi.org/10.1177/1090198106297855}.
\newblock Publisher: SAGE Publications Inc.

\bibitem[Martin and Sharma(2022)]{martin_getting_2022}
Justin~D. Martin and Krishna Sharma.
\newblock Getting news from social media influencers and from digital legacy news outlets and print legacy news outlets in seven countries: {The} “more-and-more” phenomenon and the new opinion leadership.
\newblock \emph{Newspaper Research Journal}, 43\penalty0 (3):\penalty0 276--299, September 2022.
\newblock ISSN 0739-5329.
\newblock \doi{10.1177/07395329221105507}.
\newblock URL \url{https://doi.org/10.1177/07395329221105507}.
\newblock Publisher: SAGE Publications Inc.

\bibitem[Oueslati et~al.(2023)Oueslati, Mejri, Alotaibi, and Ayouni]{oueslati_recognition_2023}
Wided Oueslati, Siwar Mejri, Shaha Alotaibi, and Sarra Ayouni.
\newblock Recognition of {Opinion} {Leaders} in {Social} {Networks} {Using} {Text} {Posts}’ {Trajectory} {Scoring} and {Users}’ {Comments} {Sentiment} {Analysis}.
\newblock \emph{IEEE Access}, PP:\penalty0 1--1, January 2023.
\newblock \doi{10.1109/ACCESS.2023.3329049}.

\bibitem[Emmons(2020)]{emmons_joy_2020}
Robert~A. Emmons.
\newblock Joy: {An} introduction to this special issue.
\newblock \emph{The Journal of Positive Psychology}, 15\penalty0 (1):\penalty0 1--4, January 2020.
\newblock ISSN 1743-9760.
\newblock \doi{10.1080/17439760.2019.1685580}.
\newblock URL \url{https://doi.org/10.1080/17439760.2019.1685580}.
\newblock Publisher: Routledge.

\bibitem[Cannon(1927)]{cannon_james-lange_1927}
Walter~B. Cannon.
\newblock The {James}-{Lange} {Theory} of {Emotions}: {A} {Critical} {Examination} and an {Alternative} {Theory}.
\newblock \emph{The American Journal of Psychology}, 39\penalty0 (1/4):\penalty0 106--124, 1927.
\newblock ISSN 00029556.
\newblock \doi{10.2307/1415404}.
\newblock URL \url{http://www.jstor.org/stable/1415404}.
\newblock Publisher: University of Illinois Press.

\bibitem[Zhan et~al.(2015)Zhan, Ren, Fan, and Luo]{zhan_distinctive_2015}
Jun Zhan, Jun Ren, Jin Fan, and Jing Luo.
\newblock Distinctive effects of fear and sadness induction on anger and aggressive behavior.
\newblock \emph{Front. Psychol.}, 6, June 2015.
\newblock ISSN 1664-1078.
\newblock \doi{10.3389/fpsyg.2015.00725}.
\newblock URL \url{http://dx.doi.org/10.3389/fpsyg.2015.00725}.
\newblock Publisher: Frontiers Media SA.

\bibitem[Russell and Giner-Sorolla(2013)]{russell_bodily_2013}
Pascale~Sophie Russell and Roger Giner-Sorolla.
\newblock Bodily moral disgust: what it is, how it is different from anger, and why it is an unreasoned emotion.
\newblock \emph{Psychol. Bull.}, 139\penalty0 (2):\penalty0 328--351, March 2013.
\newblock ISSN 0033-2909.
\newblock \doi{10.1037/a0029319}.
\newblock URL \url{http://dx.doi.org/10.1037/a0029319}.
\newblock Publisher: American Psychological Association (APA).

\bibitem[Hutcherson and Gross(2011)]{hutcherson_moral_2011}
Cendri~A Hutcherson and James~J Gross.
\newblock The moral emotions: a social-functionalist account of anger, disgust, and contempt.
\newblock \emph{J. Pers. Soc. Psychol.}, 100\penalty0 (4):\penalty0 719--737, April 2011.
\newblock ISSN 0022-3514.
\newblock \doi{10.1037/a0022408}.
\newblock URL \url{http://dx.doi.org/10.1037/a0022408}.
\newblock Publisher: American Psychological Association (APA).

\bibitem[Salerno and Peter-Hagene(2013)]{salerno_interactive_2013}
Jessica~M. Salerno and Liana~C. Peter-Hagene.
\newblock The interactive effect of anger and disgust on moral outrage and judgments.
\newblock \emph{Psychological Science}, 24\penalty0 (10):\penalty0 2069--2078, 2013.
\newblock ISSN 1467-9280(Electronic),0956-7976(Print).
\newblock \doi{10.1177/0956797613486988}.
\newblock Place: US Publisher: Sage Publications.

\bibitem[Butz et~al.(2007)Butz, Sigaud, Pezzulo, and Baldassarre]{butz_anticipatory_2007}
Martin~V. Butz, Olivier Sigaud, Giovanni Pezzulo, and Gianluca Baldassarre, editors.
\newblock \emph{Anticipatory {Behavior} in {Adaptive} {Learning} {Systems}: {From} {Brains} to {Individual} and {Social} {Behavior}}.
\newblock Springer Berlin Heidelberg, Berlin, Heidelberg, 2007.
\newblock ISBN 978-3-540-74262-3.

\bibitem[Butz and Kutter(2017)]{butz_behavioral_2017}
Martin~V. Butz and Esther~F. Kutter.
\newblock Behavioral {Flexibility} and {Anticipatory} {Behavior}.
\newblock In Martin~V. Butz and Esther~F. Kutter, editors, \emph{How the {Mind} {Comes} into {Being}: {Introducing} {Cognitive} {Science} from a {Functional} and {Computational} {Perspective}}, pages 131--154. Oxford University Press, Oxford, January 2017.
\newblock ISBN 978-0-19-873969-2.
\newblock \doi{10.1093/acprof:oso/9780198739692.003.0006}.
\newblock URL \url{https://doi.org/10.1093/acprof:oso/9780198739692.003.0006}.

\bibitem[Hoffmann(2003)]{hoffmann_anticipatory_2003}
Joachim Hoffmann.
\newblock Anticipatory {Behavioral} {Control}.
\newblock In Martin~V. Butz, Olivier Sigaud, and Pierre Gérard, editors, \emph{Anticipatory {Behavior} in {Adaptive} {Learning} {Systems}: {Foundations}, {Theories}, and {Systems}}, pages 44--65. Springer Berlin Heidelberg, Berlin, Heidelberg, 2003.
\newblock ISBN 978-3-540-45002-3.
\newblock \doi{10.1007/978-3-540-45002-3_4}.
\newblock URL \url{https://doi.org/10.1007/978-3-540-45002-3_4}.

\bibitem[Odou and Schill(2020)]{odou_how_2020}
Philippe Odou and Marie Schill.
\newblock How anticipated emotions shape behavioral intentions to fight climate change.
\newblock \emph{Journal of Business Research}, 121:\penalty0 243--253, December 2020.
\newblock ISSN 0148-2963.
\newblock \doi{10.1016/j.jbusres.2020.08.047}.
\newblock URL \url{https://www.sciencedirect.com/science/article/pii/S0148296320305518}.

\bibitem[Pleeging et~al.(2022)Pleeging, van Exel, and Burger]{pleeging_characterizing_2022}
Emma Pleeging, Job van Exel, and Martijn Burger.
\newblock Characterizing {\textit{Hope}}: {An} {Interdisciplinary} {Overview} of the {Characteristics} of {\textit{Hope}}.
\newblock \emph{Applied Research in Quality of Life}, 17\penalty0 (3):\penalty0 1681--1723, June 2022.
\newblock ISSN 1871-2576.
\newblock \doi{10.1007/s11482-021-09967-x}.
\newblock URL \url{https://doi.org/10.1007/s11482-021-09967-x}.

\bibitem[Pleeging et~al.(2021)Pleeging, Burger, and van Exel]{pleeging_relations_2021}
Emma Pleeging, Martijn Burger, and Job van Exel.
\newblock The {Relations} between {\textit{Hope}} and {Subjective} {Well}-{Being}: a {Literature} {Overview} and {Empirical} {Analysis}.
\newblock \emph{Applied Research in Quality of Life}, 16\penalty0 (3):\penalty0 1019--1041, June 2021.
\newblock ISSN 1871-2576.
\newblock \doi{10.1007/s11482-019-09802-4}.
\newblock URL \url{https://doi.org/10.1007/s11482-019-09802-4}.

\bibitem[Cerreia-Vioglio et~al.(2022)Cerreia-Vioglio, Hansen, Maccheroni, and Marinacci]{cerreia-vioglio_making_2022}
Simone Cerreia-Vioglio, Lars~Peter Hansen, Fabio Maccheroni, and Massimo Marinacci.
\newblock Making {Decisions} under {Model} {Misspecification}.
\newblock \emph{arXiv [econ.TH]}, 2022.
\newblock URL \url{http://arxiv.org/abs/2008.01071}.

\bibitem[Barnett et~al.(2022)Barnett, Brock, and Hansen]{barnett_climate_2022}
Michael Barnett, William Brock, and Lars~Peter Hansen.
\newblock Climate {Change} {Uncertainty} {Spillover} in the {Macroeconomy}.
\newblock \emph{NBER Macroeconomics Annual}, 36:\penalty0 253--320, May 2022.
\newblock ISSN 0889-3365.
\newblock \doi{10.1086/718668}.
\newblock URL \url{https://doi.org/10.1086/718668}.
\newblock Publisher: The University of Chicago Press.

\bibitem[Spearman(1904)]{spearman_proof_1904}
Charles Spearman.
\newblock The proof and measurement of association between two things.
\newblock \emph{The American Journal of Psychology}, 15\penalty0 (1):\penalty0 72--101, 1904.
\newblock \doi{10.2307/1412159}.
\newblock URL \url{https://doi.org/10.2307/1412159}.

\bibitem[Zar(1999)]{zar_biostatistical_1999}
Jerrold H. Zar.
\newblock \emph{Biostatistical Analysis}.
\newblock 4th Edition, Prentice Hall, Upper Saddle River, 1999.

\bibitem[Kurth(2015)]{kurth_moral_2015}
Charlie Kurth.
\newblock Moral anxiety and moral agency.
\newblock 2015.

\bibitem[Kurth(2018)]{kurth_anxious_2018}
Charlie Kurth.
\newblock \emph{The {Anxious} {Mind}: {An} {Investigation} into the {Varieties} and {Virtues} of {Anxiety}}.
\newblock The MIT Press, April 2018.
\newblock ISBN 978-0-262-34549-1.
\newblock \doi{10.7551/mitpress/11168.001.0001}.
\newblock URL \url{https://doi.org/10.7551/mitpress/11168.001.0001}.

\bibitem[Liu et~al.(2024)Liu, Jiao, Zhou, Kendrick, Yao, Gong, Xiang, Jia, Zhang, Zhang, Feng, and Becker]{liu_neural_2024}
Xiqin Liu, Guojuan Jiao, Feng Zhou, Keith~M. Kendrick, Dezhong Yao, Qiyong Gong, Shitong Xiang, Tianye Jia, Xiao-Yong Zhang, Jie Zhang, Jianfeng Feng, and Benjamin Becker.
\newblock A neural signature for the subjective experience of threat anticipation under uncertainty.
\newblock \emph{Nature Communications}, 15\penalty0 (1):\penalty0 1544, February 2024.
\newblock ISSN 2041-1723.
\newblock \doi{10.1038/s41467-024-45433-6}.
\newblock URL \url{https://doi.org/10.1038/s41467-024-45433-6}.

\bibitem[Königs(2022)]{konigs_what_2022}
Peter Königs.
\newblock What is {Techno}-{Optimism}?
\newblock \emph{Philosophy \& Technology}, 35\penalty0 (3):\penalty0 63, July 2022.
\newblock ISSN 2210-5441.
\newblock \doi{10.1007/s13347-022-00555-x}.
\newblock URL \url{https://doi.org/10.1007/s13347-022-00555-x}.\\


\end{thebibliography}

\section{Declarations}

\paragraph{Data availability.}\label{OSF_repo}
The complete dataset and the results generated and/or analysed during this study are available in the OSF repository, and can be accessed via the following link: \url{https://osf.io/z925y}.\\

\paragraph{Competing Interests.}
The authors declare no potential conflict of interests.\\

\paragraph{Ethical approval.}
Ethical approval was not required for this study, as it  focused exclusively on public figures (key opinion leaders) who share content professionally with the expectation of public engagement

\paragraph{Informed Consent.}
This article does not contain any studies with human participants performed by any of the authors.\\

\paragraph{Author contributions.}
The manuscript represents a collaborative effort by all authors. The second author conceptualized and inspired the entire research direction, and designed the methodological approach. The first author collected data, conducted BERTopic analysis, performed statistical analysis of emotions, attitude and sentiment, prepared a data and code repository, and provided funding for open-access publication. The third author contributed the sentiment analysis model and provided GPU hardware. Following thorough critical revisions and editing, all authors reviewed and approved the final manuscript for publication.\\


\listoftables
\listoffigures

\end{document}